\documentclass[prd,twocolumn]{revtex4-2}
\usepackage{amsmath}
\usepackage{amssymb}
\usepackage[scr=boondox]{mathalpha}
\usepackage{graphicx}

\usepackage{tabularx}
\usepackage[table]{xcolor}
\usepackage{booktabs}

\usepackage{hyperref}

\allowdisplaybreaks

\begin{document}

\title{Particle trajectories in light pulse spacetime}

\author{Riccardo Falcone}
\affiliation{Department of Physics, University of Sapienza, Piazzale Aldo Moro 5, 00185 Rome, Italy}

\author{Claudio Conti}
\affiliation{Department of Physics, University of Sapienza, Piazzale Aldo Moro 5, 00185 Rome, Italy}

\begin{abstract}
In our previous work (Phys.\ Rev.\ Research 7, 033079), we derived the metric tensor for cylindrically shaped pulses with uniform energy density. Building upon that framework, we derive the complete set of geodesics with zero angular velocity. We show that perturbations in particle trajectories may be observed in gamma ray bursts. Also, deviations in the motion of moving particles are significantly larger than those previously found for particles that are initially at rest.
\end{abstract}

\maketitle

\section{Introduction}

The gravitational effects of electromagnetic radiation on test particles have been extensively studied under the assumption of weak gravitational field and by using the retarded integral approach \cite{PhysRev.37.602, Hegarty1969, Rätzel_2016, PhysRevD.95.084008, Kadlecova2017}. In their seminal work, Tolman, Ehrenfest, and Podolsky \cite{PhysRev.37.602} analyzed the weak gravitational field generated by steady beams and passing pulses of light. By solving the corresponding geodesic equation, they examined the acceleration of test particles. A particle positioned equidistant from the two ends of the beam experiences no net parallel acceleration; on the contrary, a transverse component of the acceleration is predicted. Additionally, test light rays propagating in the beam direction do not experience gravitational effects.

Hegarty \cite{Hegarty1969} extended this analysis by considering the emitter and absorber of the radiation. He also argued that the gravitational effects of the light pulse arise solely from its emission and absorption. Rätzel \textit{et al.}~\cite{Rätzel_2016} further showed that these gravitational effects are confined to spherical shells expanding at the speed of light, representing the spacetime imprints of the emission and absorption events.  Near the light beam and far from the emitter and absorber, the deflection of test particles is predominantly transverse to the trajectory of the pulse. The emission event of the pulse causes an attractive effect, while the absorption event produces a repulsive effect.

The trajectories of particles in the spacetime of a light pulse have also been derived in the exact theory of general relativity \cite{Bonnor1969, https://doi.org/10.1002/prop.201000088, PhysRevD.89.104049, ylvn-3ybm}, thereby overcoming the approximations of the retarded integral approach. Bonnor's \cite{Bonnor1969} exact solution to the Einstein equation describes a null fluid traveling along an infinite straight path, accompanied by a gravitational wave (GW) propagating in the same direction as the light pulse. By solving the geodesic equations, Bonnor examined the trajectories of particles in this spacetime and proved that, in general, only massless particles propagating in the same direction as the beam remain undeflected. Van Holten \cite{https://doi.org/10.1002/prop.201000088} and Bini \textit{et al.}~\cite{PhysRevD.89.104049} studied the geodesics in the spacetime of an electromagnetic plane wave, by using the Brinkmann and the Rosen coordinates, respectively. Additionally, Bini \textit{et al.}~explored the accelerated orbits of charged particles undergoing inverse Compton scattering, as well as trajectories of particles with spin.

In our previous work \cite{ylvn-3ybm}, we derived the exact solution to the Einstein equations for cylindrically shaped pulses with constant energy density, assuming the pulse is emitted from infinity. This simplified model provides an effective local description of any pulse at large distances from its source. In line with Refs.\ \cite{Hegarty1969, Bonnor2009, Rätzel_2016}, the pulse is accompanied by a GW that shares the same plane wavefront as the pulse. We derived the geodesics for a particle initially at rest and subsequently hit by the GW.

In this paper, we extend our analysis to derive the trajectories of particles with arbitrary initial conditions, including those crossing the pulse region and those with nonzero initial velocities. Our focus is on geodesics with zero angular velocity. We solve the geodesic equations within each of the three spacetime regions (pulse, GW and exterior), then match the solutions at each boundary to derive a comprehensive description of particle trajectories throughout the entire spacetime. We recover the results of previous studies \cite{PhysRev.37.602, Rätzel_2016, Bonnor1969} showing that massless particles copropagating with the beam remain undeflected, while the deflection of massive particles in the weak gravity regime is primarily transverse to the trajectory of the beam. Additionally, we demonstrate that for particles with nonzero initial velocity, the perturbations induced by the GW are of the order of $\epsilon \log(\epsilon)$, in contrast to $\epsilon$-order perturbations found for particles initially at rest \cite{ylvn-3ybm}. Here, $\epsilon$ is the dimensionless parameter assumed to be small in the weak gravity limit.

The paper is organized as follows. In Sec.\ \ref{Spacetime_of_a_single_pulse}, we provide a brief overview of the pulse spacetime, following the results of our previous work \cite{ylvn-3ybm}. In Sec.\ \ref{Local_geodesics}, we derive the geodesics in the pulse, GW, and exterior regions. Continuity conditions at the boundaries are addressed in Sec.\ \ref{Matching_conditions}. In Sec.\ \ref{Geodesics}, we combine these results to derive the particle trajectories perturbed by the passage of the pulse or the GW. Conclusions are drawn in Sec.\ \ref{Conclusions}.

\section{Spacetime of a pulse}\label{Spacetime_of_a_single_pulse}

\begin{figure}[h]
\includegraphics[]{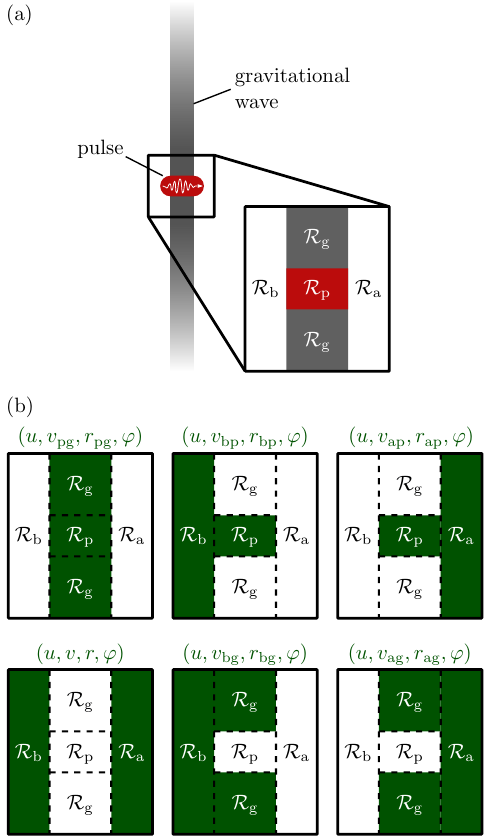}
\caption{Electromagnetic pulse propagating alongside a gravitational wave (GW). (a) The spacetime is divided into four distinct regions: $\mathcal{R}_\text{p}$, $\mathcal{R}_\text{g}$, $\mathcal{R}_\text{a}$ and $\mathcal{R}_\text{b}$. The regions $\mathcal{R}_\text{p}$ and $\mathcal{R}_\text{g}$ correspond to the pulse and the GW, respectively, while $\mathcal{R}_\text{a}$ and $\mathcal{R}_\text{b}$ represent the Minkowski spacetime before and after the passage of the pulse and the GW. (b) Various coordinate systems are used to describe the spacetime, each covering different combinations of regions. The coordinates $(u, v_{\text{p}\text{g}}, r_{\text{p}\text{g}}, \varphi)$ are used for the interior regions $\mathcal{R}_\text{p}$ and $\mathcal{R}_\text{g}$, while $(u, v, r, \varphi)$ are the standard Minkowski coordinates for $\mathcal{R}_\text{a} \cup \mathcal{R}_\text{b}$. Additionally, for each $\xi \in \{ \text{a}, \text{b} \}$ and $\eta \in \{ \text{p}, \text{g} \}$, the coordinate system $(u, v_{\xi \eta}, r_{\xi \eta}, \varphi)$ spans the region $\mathcal{R}_\xi \cup \mathcal{R}_\eta$. }\label{Spacetime_figure}
\end{figure}

Throughout this paper, we consider the spacetime associated with a light pulse propagating from infinity. In this section, we provide a brief overview of this spacetime. We use the results of our previous work \cite{ylvn-3ybm}, where we solved the Einstein equation for a cylindrically-shaped uniform energy density propagating at the speed of light. The pulse is accompanied by a gravitational shock wave propagating in the same direction and with a plane wavefront. The global structure of this spacetime is illustrated in Fig.\ \ref{Spacetime_figure}.

We denote the spacetime regions before and after the passage of the pulse and the GW as $\mathcal{R}_\text{a}$ and $\mathcal{R}_\text{b}$. These regions are described through the Minkowski coordinate system $(u, v, r, \varphi)$, with the metric given by
\begin{subequations}
\begin{align}
& \left.  ds^2 \right|_{\mathcal{R}_\text{a}} = -2dudv + dr^2 +r^2 d\varphi^2, \\
 & \left.  ds^2 \right|_{\mathcal{R}_\text{b}} = -2dudv + dr^2 +r^2 d\varphi^2.
\end{align}
\end{subequations}
Here, $u=(ct-z)/\sqrt{2}$ and $v=(ct+z)/\sqrt{2}$ represent the co-propagating and counter-propagating light-cone coordinates, $t$ is the time coordinate, $z$ is the direction of propagation, $r \in [0, \infty)$ is the radial coordinate perpendicular to $z$, $\varphi\in[0,2\pi)$ is the azimuthal coordinate and $c$ is the speed of light. In this coordinate system, the regions $\mathcal{R}_\text{a}$ and $\mathcal{R}_\text{b}$ are defined by $u<\bar{u}_\text{a}$ and $u>\bar{u}_\text{b}$, respectively, where $\bar{u}_\text{a}$ and $\bar{u}_\text{b}$ mark the boundaries of the pulse and the GW in the copropagating coordinate $u$.

The spacetime regions corresponding to the pulse ($\mathcal{R}_\text{p}$) and the GW ($\mathcal{R}_\text{g}$)  are described via the coordinates $(u, v_{\text{p}\text{g}}, r_{\text{p}\text{g}}, \varphi)$ and the metric
\begin{equation}\label{g_vr_pg}
\left.  ds^2 \right|_{\mathcal{R}_\text{p} \cup \mathcal{R}_\text{g}} = - \Phi(r_{\text{p}\text{g}}) du^2 -2dudv_{\text{p}\text{g}} + dr_{\text{p}\text{g}}^2 +r_{\text{p}\text{g}}^2 d\varphi^2.
\end{equation}
The function $\Phi(r_{\text{p}\text{g}})$ is defined as
\begin{equation}\label{Phi}
\Phi(r_{\text{p}\text{g}}) = \begin{cases}
\epsilon (r_{\text{p}\text{g}}/\bar{r}_\text{p})^2 & \text{if } r_{\text{p}\text{g}} \leq \bar{r}_\text{p},  \\
\epsilon \left[1 + 2\log(r_{\text{p}\text{g}}/\bar{r}_\text{p})  \right] & \text{if } r_{\text{p}\text{g}} > \bar{r}_\text{p},
\end{cases}
\end{equation}
with $\bar{r}_\text{p}$ as the radius of the pulse, $\epsilon = \kappa I \bar{r}_\text{p}^2 /c$ as the coupling between the pulse and the gravitational field and $I$ as the intensity of the pulse. In these coordinates, the stress-energy tensor is defined as $T = \theta(\bar{r}_\text{p} - r_{\text{p}\text{g}})  ( 2 I /c) \partial_{v_{\text{p}\text{g}}} \otimes \partial_{v_{\text{p}\text{g}}}$, with $\theta (\cdot)$ as the Heaviside step function.

To connect the interior regions ($\mathcal{R}_\text{p}$ and $\mathcal{R}_\text{g}$) with the exterior regions ($\mathcal{R}_\text{a}$ and $\mathcal{R}_\text{b}$), we introduce additional coordinate systems $(u, v_{\xi \eta}, r_{\xi \eta}, \varphi)$, with $\xi \in \{ \text{a}, \text{b} \}$ and $\eta \in \{ \text{p}, \text{g} \}$. Within the region $\mathcal{R}_\eta$, the coordinates $(u, v_{\xi \eta}, r_{\xi \eta}, \varphi)$ are defined through the transformation
\begin{align}\label{vr_pg_vr_xieta}
& v_{\text{p}\text{g}} = v_{\xi \eta} + V_\eta \left(u-\bar{u}_\xi ,r_{\xi\eta} \right), && r_{\text{p}\text{g}} = R_\eta \left(u-\bar{u}_\xi ,r_{\xi\eta} \right),
\end{align}
with
\begin{subequations}\label{VR_eta}
\begin{align}
& V_{\text{p}}(u,r) = \frac{R_{\text{p}}(u,r) \dot{R}_{\text{p}}(u,r)}{2}, \label{V_p} \\
& R_{\text{p}}(u,r) = \cos \left( \sqrt{\epsilon} \frac{u}{\bar{r}_{\text{p}}} \right) r, \label{R_p}   \\
& V_{\text{g}}(u, r) =  R_{\text{g}}(u,r) \dot{R}_{\text{g}}(u,r) + \frac{\epsilon}{2}  \left[1-2 \log \left(\frac{r}{\bar{r }_{\text{p}}}\right)\right] u,  \label{V_g}  \\
& R_{\text{g}}(u, r) = \exp \left\lbrace-\left[\text{erf}^{-1}\left(\sqrt{\frac{2 \epsilon }{\pi }} \frac{u }{r }\right)\right]^2\right\rbrace r,   \label{R_g} \\
& \dot{R}_\eta(u,r) = \partial_u R_\eta(u,r),
\end{align}
\end{subequations}
and $\text{erf}^{-1}(\cdot)$ as the inverse error function. These coordinates cover the subregions $\mathcal{D}_\xi \subset \mathcal{R}_\text{g}$ defined as $| u - \bar{u}_\xi| / r_{\xi\text{g}} < \sqrt{\pi/2 \epsilon}$. In terms of $(u, v_{\xi \eta}, r_{\xi \eta}, \varphi)$, the metric within the pulse and the GW is expressed as
\begin{align}\label{g_p_vr_xieta}
\left.  ds^2 \right|_{\mathcal{R}_\eta}  = & - 2 du dv_{\xi\eta} + \left[ R'_\eta(u- \bar{u}_\xi, r_{\xi\eta}) \right]^2 dr_{\xi\eta}^2 \nonumber \\
& +  R_\eta^2 (u- \bar{u}_\xi, r_{\xi\eta}) d\varphi^2,
\end{align}
with $R'_\eta(u,r) = \partial_r R_\eta(u,r)$. This coordinate system can be smoothly extended into the exterior region $\mathcal{R}_\xi$ by assuming $v_{\xi \eta} = v$ and $r_{\xi\eta} = r$ when restricted to $\mathcal{R}_\xi$.

\section{Local geodesics}\label{Local_geodesics}

In Sec.\ \ref{Spacetime_of_a_single_pulse} we introduced the spacetime associated to a cylindrical light pulse. In this section, we use this framework to examine the trajectories of particles within each region ($\mathcal{R}_\text{p}$, $\mathcal{R}_\text{g}$, $\mathcal{R}_\text{a}$ and $\mathcal{R}_\text{b}$). Specifically, we solve the geodesic equation 
\begin{equation}\label{geodesic_equation}
\frac{d^2 x^\rho}{ds^2}+ \Gamma^\rho{}_{\mu\nu} \frac{d x^\mu}{ds} \frac{d x^\nu}{ds}  = 0
\end{equation}
in each region. Here, $\Gamma^\rho{}_{\mu\nu} $ are the Christoffel symbols associated to the metric $g_{\mu\nu}$.

\subsection{Pulse region ($\mathcal{R}_\text{p}$)}\label{Light_pulse}

In this subsection, we focus on the pulse region $\mathcal{R}_\text{p}$. By using Eq.\ (\ref{g_vr_pg}) with $\Phi(r_{\text{p}\text{g}}) =
\epsilon (r_{\text{p}\text{g}}/\bar{r}_\text{p})^2$, we solve the geodesic equation (\ref{geodesic_equation}) for a particle moving within $\mathcal{R}_\text{p}$. Here, the variables $(x^0(s), x^1(s), x^2(s), x^3(s)) = (u(s), v_{\text{p}\text{g}}(s), r_{\text{p}\text{g}}(s), \varphi(s))$ correspond to the $(u, v_{\text{p}\text{g}}, r_{\text{p}\text{g}}, \varphi)$ coordinates of the trajectory of the particle, parametrized by $s$. In this case, Eq.\ (\ref{geodesic_equation}) reduces to
\begin{subequations}\label{geodesic_equation_p_pg}
\begin{align}
& u''(s) = 0, \label{geodesic_equation_p_pg_a} \\
& v_{\text{pg}}''(s) = - 2 \epsilon \frac{u'(s) r_{\text{pg}}(s)  r_{\text{pg}}'(s)}{\bar{r}_{\text{p}}^2} , \label{geodesic_equation_p_pg_b} \\
& r_{\text{pg}}''(s) = \left\lbrace  - \epsilon  \frac{[u'(s)]^2}{\bar{r}_{\text{p}}^2} + [\varphi '(s)]^2\right\rbrace r_{\text{pg}}(s), \label{geodesic_equation_p_pg_c}\\
& \varphi ''(s) =- 2 \frac{ r_{\text{pg}}'(s) \varphi '(s) }{r_{\text{pg}}(s)}. \label{geodesic_equation_p_pg_d}
\end{align}
\end{subequations}

We start by considering the case where $u(s)$ is constant (i.e., $u(s) = \bar{u}_\gamma$, for some $\bar{u}_\gamma \in [\bar{u}_\text{a}, \bar{u}_\text{b}]$), which trivially satisfies Eq.\ (\ref{geodesic_equation_p_pg_a}). Under this assumption, the magnitude of the four-velocity $U^\mu = dx^\mu/ds$ is $ U^\mu(s) U_\mu(s) =[r_{\text{p}\text{g} }'(s)]^2 + r_{\text{p}\text{g}}^2(s) [\varphi'(s)]^2$. The condition of no superluminality $U^\mu(s) U_\mu(s) \leq 0$ implies that $r_{\text{p}\text{g}}(s) $ must be constant (i.e., $r_{\text{p}\text{g}}(s) = \bar{r}_\gamma$, for some $\bar{r}_\gamma \in [0,\bar{r}_\text{p}]$). Additionally, either $\bar{r}_\gamma = 0$, or $\varphi(s)$ must also be constant (i.e., $\varphi(s) = \bar{\varphi}_\gamma$). In both cases, we find that $ U^\mu(s) U_\mu(s)=0$, indicating that the particle is massless. When $u(s) = \bar{u}_\gamma$, Eq.\ (\ref{geodesic_equation_p_pg_b}) leads to $v_{\text{p} \text{g}}''(s) = 0$, which means that $v_{\text{p} \text{g}}(s) $ is linear in $s$. We choose an affine parametrization such that $v_{\text{p} \text{g}}(s) = s$. As a result, we find that the general solution to Eq.\ (\ref{geodesic_equation_p_pg}) for constant $u(s)$ is
\begin{equation}\label{geodesic_u_gamma}
(u(s), v_{\text{p}\text{g}}(s), r_{\text{p}\text{g}}(s), \varphi(s)) = (\bar{u}_\gamma, s, \bar{r}_\gamma, \bar{\varphi}_\gamma).
\end{equation}
Equation (\ref{geodesic_u_gamma}) describes the trajectory of a massless particle traveling in the same direction as the pulse. This result is consistent with the findings of Tolman \textit{et al.}~\cite{PhysRev.37.602} and Bonnor \cite{Bonnor1969}, who showed that massless particles copropagating with the pulse remain undeflected.

When $u(s)$ is not constant, Eq.\ (\ref{geodesic_equation_p_pg_a}) implies that $u(s)$ must be linear in $s$. By translating the parameter $s$, we can write $u(s) $ as $u(s) = c_{\text{p},1} s$, with $c_{\text{p},1}$ as a nonzero constant. In this case, Eqs.\ (\ref{geodesic_equation_p_pg_b}) and (\ref{geodesic_equation_p_pg_c}) lead to
\begin{subequations}\label{geodesic_equation_p_pg_2}
\begin{align}
& v_{\text{pg}}''(s) =  - 2 \epsilon c_{\text{p},1} \frac{r_{\text{pg}}(s)  r_{\text{pg}}'(s)}{\bar{r}_{\text{p}}^2},\\
 & r_{\text{pg}}''(s) = \left\lbrace  -   \frac{\epsilon c_{\text{p},1}^2}{\bar{r}_{\text{p}}^2} + [\varphi '(s)]^2\right\rbrace r_{\text{pg}}(s),
\end{align}
\end{subequations}
whereas Eq.\ (\ref{geodesic_equation_p_pg_d}) is solved by
\begin{equation}
\varphi(s) = c_{\text{p},6} + c_{\text{p},7} \int_0^s \frac{ds'}{r_{\text{pg}}^2(s')} ,
\end{equation}
with $c_{\text{p},6}$ and $c_{\text{p},7}$ as constants.

\begin{figure}[p]
\includegraphics[]{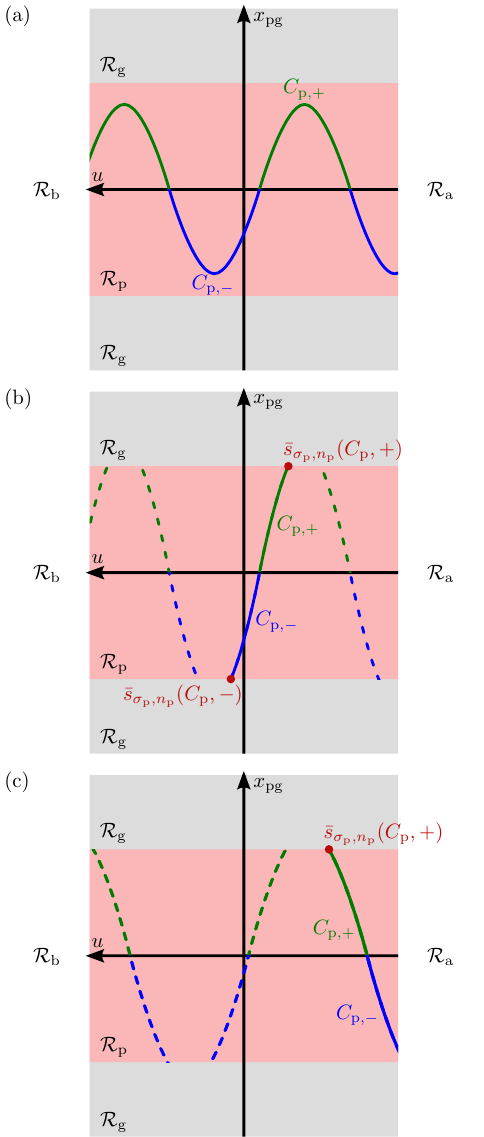}
\caption{Geodesics in the pulse region $\mathcal{R}_\text{p}$ confined to the $y_\text{pg} = 0$ plane. (a) Geodesics that do not cross the boundary $r=\bar{r}_\text{p}$ between $\mathcal{R}_\text{p}$ and $\mathcal{R}_\text{g}$. (b) Geodesics that cross the boundary $r=\bar{r}_\text{p}$ twice, at $s = \bar{s}_{\sigma_\text{p}, n_\text{p}} (C_\text{p},-)$ and $s = \bar{s}_{\sigma_\text{p}, n_\text{p}} (C_\text{p},+)$. (c) Geodesics with a single crossing point at $r=\bar{r}_\text{p}$, as $s = \bar{s}_{\sigma_\text{p}, n_\text{p}} (C_\text{p},-)$ lies beyond the boundary $u = \bar{u}_\text{a}$. The branches $C_{\text{p},-}$ and $C_{\text{p},+}$ are distinguished by $x_\text{pg} < 0$ and $x_\text{pg} > 0$, respectively. In (b) and (c), the branches $C_{\text{p},+}$ and $C_{\text{p},-}$ are further divided into separate continuous geodesics identified by the parameters $(\sigma_\text{p}, n_\text{p})$.}\label{Pulse_figure}
\end{figure}

The parameter $c_{\text{p},7}$ is associated to the tangential velocity of the particle, as the fourth component of $U^\mu$ is $c_{\text{p},7} r_{\text{pg}}^{-2}(s)$. If the tangential velocity is zero at any point, the trajectory of the particle remains confined to the plane $\varphi = c_{\text{p},6}$. In that case, Eq.\ (\ref{geodesic_equation_p_pg_2}) reduces to
\begin{align}
& v_{\text{pg}}''(s) =  - 2 \epsilon c_{\text{p},1} \frac{r_{\text{pg}}(s)  r_{\text{pg}}'(s)}{\bar{r}_{\text{p}}^2},
 & r_{\text{pg}}''(s) =  -  \epsilon c_{\text{p},1}^2 \frac{r_{\text{pg}}(s)}{\bar{r}_{\text{p}}^2} .
\end{align}
The solutions to these equations are
\begin{subequations}
\begin{align}
& v_{\text{pg}}(s) = c_{\text{p},2} s + c_{\text{p},3} +V_\text{p} \left( c_{\text{p},1} s + c_{\text{p},4}, c_{\text{p},5} \right), \\
& r_{\text{pg}}(s)  = R_\text{p} \left( c_{\text{p},1} s + c_{\text{p},4}, c_{\text{p},5}\right),
\end{align}
\end{subequations}
where the functions $V_{\text{p}}(u,r)$ and $R_\text{p}(u,r)$ are defined by Eqs.\ (\ref{V_p}) and (\ref{R_p}), respectively. As a result, we find that the general solution to Eq.\ (\ref{geodesic_equation_p_pg}) for nonconstant $u(s)$ and vanishing tangential velocity is
\begin{align}\label{geodesics_p}
& \left. (u(s), v_{\text{p}\text{g}}(s), r_{\text{p}\text{g}}(s), \varphi(s)) \right|_{\mathcal{R}_\text{p}}  = (c_{\text{p},1} s, c_{\text{p},2} s + c_{\text{p},3} \nonumber \\
& + V_\text{p} \left( c_{\text{p},1} s + c_{\text{p},4}, c_{\text{p},5} \right), R_\text{p} \left( c_{\text{p},1} s + c_{\text{p},4}, c_{\text{p},5}\right), c_{\text{p},6}).
\end{align}
Among these solutions, those confined to the plane $y_\text{pg} = 0$ are depicted in Fig.\ \ref{Pulse_figure}. The coordinates $x_\text{pg}$, $y_\text{pg}$ and $z_\text{pg}$ are defined as $x_\text{pg} = r_\text{pg} \cos (\varphi)$, $y_\text{pg} = r_\text{pg} \sin (\varphi)$ and $z_\text{pg} = (v_\text{pg} - u)/\sqrt{2}$.

By restricting the right-hand side of Eq.\ (\ref{geodesics_p}) to the interior region of the pulse $\mathcal{R}_\text{p}$, we find that not all geodesics are uniquely determined by the parameters $(c_{\text{p},1}, \dots, c_{\text{p},6})$. In the particular case where $| c_{\text{p},5} | \leq \bar{r}_\text{p}$, each configuration $(c_{\text{p},1}, \dots, c_{\text{p},6})$ corresponds to multiple discontinuous lines that periodically cross $r_\text{pg}=0$ but never cross the boundary $r_\text{pg} = \bar{r}_\text{p}$. These segments can be smoothly connected to each other through the geodesics parameterized by $(c_{\text{p},1}, \dots, - c_{\text{p},5} , c_{\text{p},6} + \pi)$. The pair of configurations $C_{\text{p},+} = (c_{\text{p},1}, \dots, c_{\text{p},5}, c_{\text{p},6})$ and $C_{\text{p},-} = (c_{\text{p},1}, \dots, - c_{\text{p},5} , c_{\text{p},6} + \pi )$ together describe a single geodesic that remains confined to $\mathcal{R}_\text{p}$ without crossing $r_\text{pg} = \bar{r}_\text{p}$ [see Fig.\ \ref{Pulse_figure}(a)].

By restricting the parameters to $ c_{\text{p},4}\in[0, 2 \pi \bar{r}_{\text{p}} / \sqrt{\epsilon})$, $ c_{\text{p},5}\in[0, \bar{r}_\text{p}]$ and $ c_{\text{p},6}\in[0, \pi)$, a one-to-one correspondence is established between $ C_\text{p} = (c_{\text{p},1}, \dots, c_{\text{p},6})$ and the connected geodesics in $\mathcal{R}_\text{p}$ with zero tangential velocity that never cross the boundary $r_\text{pg} = \bar{r}_\text{p}$. These geodesics are described by
\begin{align}\label{geodesics_p_varsigma}
& \left. (u(s), v_{\text{p}\text{g}}(s), r_{\text{p}\text{g}}(s), \varphi(s)) \right|_{\mathcal{R}_\text{p}} \nonumber \\
  = \, &  (c_{\text{p},1} s, c_{\text{p},2} s + c_{\text{p},3}+V_\text{p} \left( c_{\text{p},1} s + c_{\text{p},4}, \varsigma_\text{p}(C_\text{p}, s) c_{\text{p},5} \right),  \nonumber \\
& R_\text{p} \left( c_{\text{p},1} s + c_{\text{p},4}, \varsigma_\text{p}(C_\text{p}, s) c_{\text{p},5}\right),  c_{\text{p},6}  + \theta(-\varsigma_\text{p}(C_\text{p}, s)) \pi) ,
\end{align}
where 
\begin{equation}\label{varsigma_C_p_s}
\varsigma_\text{p}(C_\text{p}, s)  = \text{sgn} \left[  \cos \left( \sqrt{\epsilon} \frac{ c_{\text{p},1} s + c_{\text{p},4} }{\bar{r}_\text{p}} \right) \right]
\end{equation}
determines the specific branch $C_{\text{p},\varsigma_\text{p}}$ where the particle is traveling at periodic intervals of time.

If $c_{\text{p},5} > \bar{r}_\text{p}$, the pairs $(C_{\text{p},+}, C_{\text{p},-})$ describe multiple disconnected geodesics that cross the boundary $r_\text{pg} = \bar{r}_\text{p}$  [see Figs.\ \ref{Pulse_figure}(b) and \ref{Pulse_figure}(c)]. In this case, the correspondence becomes one-to-many. To resolve this ambiguity, in addition to $ C_\text{p} = (c_{\text{p},1}, \dots, c_{\text{p},6})$, we introduce two parameters $\sigma_{\text{p}} \in \{ -1, +1 \}$ and $n_{\text{p}} \in\mathbb{Z}$, which select the configuration such that the branch $C_{\text{p},\varsigma_\text{p}}$ crosses the boundary $r_\text{pg} = \bar{r}_\text{p}$ at $s = \bar{s}_{\sigma_{\text{p}},n_{\text{p}}} (C_\text{p},  \varsigma_\text{p})$, given by
\begin{align}\label{s_bar_sigman}
\bar{s}_{\sigma_{\text{p}},n_{\text{p}}} (C_\text{p},   \varsigma_\text{p} ) = \, & - \frac{c_{\text{p},4}}{c_{\text{p},1}} + \frac{1}{\sqrt{\epsilon }} \frac{ \bar{r}_{\text{p}}  }{c_{\text{p},1}} \left[ \sigma_{\text{p}}\arccos\left(\frac{ \varsigma_\text{p} \bar{r}_{\text{p}}}{c_{\text{p},5}}\right) \right. \nonumber \\
& \left. + 2 n_{\text{p}}  \pi \right].
\end{align}

To ensure that the geodesic is (at least partially) confined within $u \in [\bar{u}_\text{a}, \bar{u}_\text{b}]$, the parameter $n_{\text{p}} $ must satisfy the constraint
\begin{subequations}
\begin{align}
& n_{\text{p}} \geq \frac{\sqrt{\epsilon }}{2 \pi }  \frac{ \bar{u}_\text{a} + c_{\text{p},4} }{\bar{r}_{\text{p}}  } - \frac{\sigma_{\text{p}}}{2\pi}\arccos\left(\frac{ \sigma_{\text{p}} \bar{r}_{\text{p}}}{ c_{\text{p},5}}\right), \\
 & n_{\text{p}} \leq   \frac{\sqrt{\epsilon }}{2 \pi }  \frac{ \bar{u}_\text{b} + c_{\text{p},4} }{\bar{r}_{\text{p}}  }  - \frac{\sigma_{\text{p}}}{2\pi}\arccos\left(- \frac{\sigma_{\text{p}} \bar{r}_{\text{p}}}{c_{\text{p},5}}\right).
\end{align}
\end{subequations}
Under this condition, the mapping between $(C_\text{p}, \sigma_\text{p}, n_\text{p})$ and the connected geodesics within $\mathcal{R}_\text{p}$ with zero tangential velocity that cross the boundary $r_\text{pg} = \bar{r}_\text{p}$ is one-to-one. Explicitly, these geodesics are described by Eq.\ (\ref{geodesics_p_varsigma}), with the parameter $s$ confined by $ \bar{s}_{\sigma_{\text{p}},n_{\text{p}}} (C_\text{p},  \sigma_{\text{p}}) \leq s \leq \bar{s}_{\sigma_{\text{p}},n_{\text{p}}} (C_\text{p}, - \sigma_{\text{p}})$.

\subsection{GW region ($\mathcal{R}_\text{g}$)}\label{Gravitational_wave}

In this subsection, we consider the GW region $\mathcal{R}_\text{g}$ described by the metric (\ref{g_vr_pg}), with $\Phi(r_{\text{p}\text{g}}) = \epsilon [1 + 2\log(r_{\text{p}\text{g}}/\bar{r}_\text{p}) ]$. The geodesic equation (\ref{geodesic_equation}) in the Brinkmann-GW coordinates $(x^0(s), x^1(s), x^2(s), x^3(s)) = (u(s), v_{\text{p}\text{g}}(s), r_{\text{p}\text{g}}(s), \varphi(s))$ simplifies to
\begin{subequations}\label{geodesic_equation_g_pg}
\begin{align}
& u''(s) = 0, \label{geodesic_equation_g_pg_a} \\
& v_{\text{pg}}''(s) = - 2 \epsilon  \frac{u'(s) r_{\text{pg}}'(s)}{r_{\text{pg}}(s)} , \label{geodesic_equation_g_pg_b} \\
& r_{\text{pg}}''(s) =  -\epsilon \frac{ \left[u'(s)\right]^2}{r_{\text{pg}}(s)} + r_{\text{pg}}(s) \left[\varphi'(s)\right]^2, \label{geodesic_equation_g_pg_c} \\
& \varphi''(s) =- 2 \frac{r_{\text{pg}}'(s) \varphi '(s)}{r_{\text{pg}}(s)}. \label{geodesic_equation_g_pg_d} 
\end{align}
\end{subequations}
To solve Eq.\ (\ref{geodesic_equation_g_pg}), we follow the approach outlined in Sec.\ \ref{Light_pulse}. Equation (\ref{geodesic_equation_g_pg_a}) leads to two cases: $u(s)$ being constant ($u(s) = \bar{u}_\gamma$) or linear in $s$ ($u(s) = c_{\text{g},1} s$). In the constant case, the trajectory corresponds to that of a massless particle traveling in the same direction as the pulse, as previously derived in Eq.\ (\ref{geodesic_u_gamma}). In the linear case, instead, the trajectory is characterized by
\begin{equation}
\varphi(s) = c_{\text{g},6} + c_{\text{g},7} \int_0^s \frac{ds'}{r_{\text{pg}}^2(s')}
\end{equation}
and by $v_{\text{pg}}$ and $r_{\text{pg}}$ satisfying the following equations
\begin{align}\label{geodesic_equation_g_pg_3}
& v_{\text{pg}}''(s) =  - 2 \epsilon  c_{\text{g},1} \frac{ r_{\text{pg}}'(s)}{r_{\text{pg}}(s)}, & r_{\text{pg}}''(s) = - \frac{\epsilon  c_{\text{g},1}^2}{r_{\text{pg}}(s)} + \frac{c_{\text{g},7}^2}{r_{\text{pg}}^3(s)}.
\end{align}

\begin{figure}[p]
\includegraphics[]{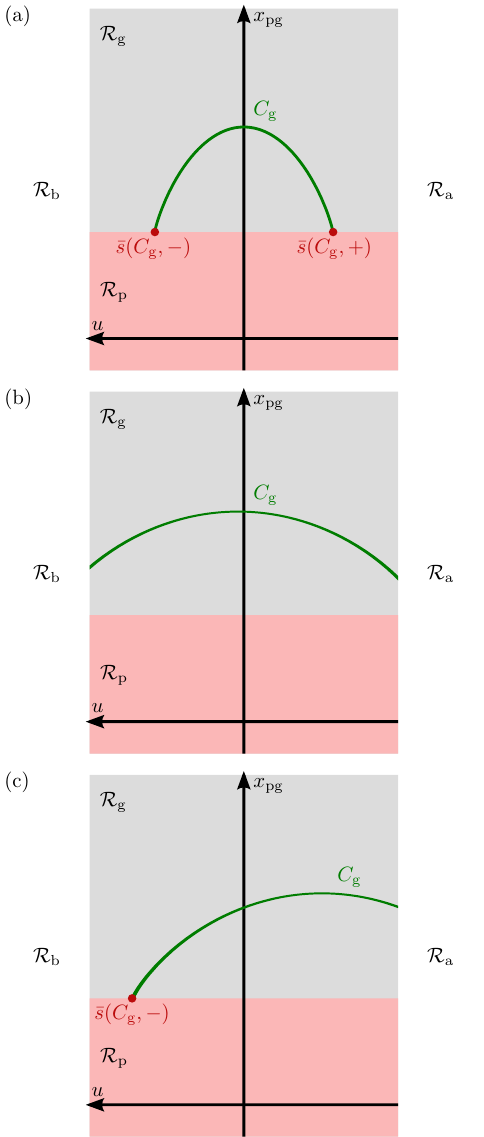}
\caption{Geodesics in the GW region $\mathcal{R}_\text{g}$ confined to the semiplane $\varphi=0$. (a) Geodesics that cross the boundary $r=\bar{r}_\text{p}$ between $\mathcal{R}_\text{p}$ and $\mathcal{R}_\text{g}$ twice, at $s = \bar{s} (C_\text{g},-)$ and $s = \bar{s} (C_\text{g},+)$. (b) Geodesics that do not cross the boundary $r=\bar{r}_\text{p}$ as $s = \bar{s} (C_\text{g},-)$ and $s = \bar{s} (C_\text{g},+)$ lie beyond the boundaries $u = \bar{u}_\text{b}$ and $u = \bar{u}_\text{a}$, respectively. (c) Geodesics with a single crossing point at $r=\bar{r}_\text{p}$, as $s = \bar{s} (C_\text{p},+)$ lies beyond the boundary $u = \bar{u}_\text{a}$.}\label{GW_figure}
\end{figure}

The fourth component of the four-velocity is $c_{\text{g},7} r_{\text{pg}}^{-2}(s)$, indicating that if the tangential velocity vanishes at any point, $c_{\text{g},7}$ must also be zero and the trajectory of the particle remains confined to the plane $\varphi = c_{\text{g},6}$. Under this condition, Eq.\ (\ref{geodesic_equation_g_pg_3}) simplifies to
\begin{align}\label{geodesic_equation_g_pg_4}
& v_{\text{pg}}''(s) =  - 2 \epsilon  c_{\text{g},1} \frac{ r_{\text{pg}}'(s)}{r_{\text{pg}}(s)}, & r_{\text{pg}}''(s) = -\frac{\epsilon  c_{\text{g},1}^2}{r_{\text{pg}}(s)}.
\end{align}
The solutions to these equations are given by
\begin{subequations}
\begin{align}
& v_{\text{pg}}(s) = c_{\text{g},2} s + c_{\text{g},3} +V_\text{g} \left( c_{\text{g},1} s + c_{\text{g},4}, c_{\text{g},5} \right), \\
& r_{\text{pg}}(s) = R_\text{g} \left( c_{\text{g},1} s + c_{\text{g},4}, c_{\text{g},5}\right),
\end{align}
\end{subequations}
where the functions $V_{\text{g}} (u,r)$ and $R_{\text{g}} (u,r)$ are defined by Eqs.\ (\ref{V_g}) and (\ref{R_g}). Thus, the general solution to Eq.\ (\ref{geodesic_equation_g_pg}) for nonconstant $u(s)$ and vanishing tangential velocity is
\begin{align}\label{geodesics_g}
& \left. (u(s), v_{\text{p}\text{g}}(s), r_{\text{p}\text{g}}(s), \varphi(s)) \right|_{\mathcal{R}_\text{g}}  = (c_{\text{g},1} s, c_{\text{g},2} s + c_{\text{g},3} \nonumber \\
& +V_\text{g} \left( c_{\text{g},1} s + c_{\text{g},4}, c_{\text{g},5} \right), R_\text{g} \left( c_{\text{g},1} s + c_{\text{g},4}, c_{\text{g},5}\right), c_{\text{g},6}).
\end{align}
The configurations with $c_{\text{g},6}=0$ (i.e., those confined to the semiplane $\varphi=0$) are shown in Fig.\ \ref{GW_figure}.

For any configuration $C_\text{g}=(c_{\text{g},1}, \dots, c_{\text{g},6})$ with $c_{\text{g},5} \geq \bar{r}_\text{p} $ and $c_{\text{g},6} \in [0, 2 \pi)$, Eq.\ (\ref{geodesics_g}) describes a single geodesic in $\mathcal{R}_\text{g}$ with zero tangential velocity. This geodesic may intersect $r_\text{pg} = \bar{r}_\text{p}$ at different values of $s$, denoted as $s = \bar{s}(C_\text{g},\varsigma_\text{g})$, with
\begin{equation}\label{s_bar_C_g_varsigma_g}
\bar{s}(C_\text{g},\varsigma_\text{g}) = -\frac{c_{\text{g},4}}{c_{\text{g},1}} -   \varsigma_\text{g} \sqrt{\frac{\pi }{2 \epsilon }} \, \text{erf}\left[\sqrt{\log \left(\frac{c_{\text{g},5}}{\bar{r}_{\text{p}}}\right)}\right] \frac{ c_{\text{g},5} }{c_{\text{g},1}},
\end{equation}
for $ \varsigma_\text{g} \in \{ -1, + 1\}$ (see Fig.\ \ref{GW_figure}).

\subsection{Exterior regions ($\mathcal{R}_\xi$)}\label{Exterior_regions}

In this subsection, we focus on the exterior regions $\mathcal{R}_\xi$, for $\xi\in\{ \text{a}, \text{b} \}$, which are described by the Minkowski coordinates $(u,v_{\xi\eta},r_{\xi\eta},\varphi)$. Here $\eta$ can take any value between p and g.

In Minkowski spacetime, particles move along straight-line trajectories. Specifically, particles with zero tangential velocity follow geodesics given by
\begin{align}\label{geodesics_xieta}
& \left. (u(s),v_{\xi\text{p}}(s),r_{\xi\text{p}}(s),\varphi(s)) \right|_{\mathcal{R}_\xi} = \left( c_{\xi,1} s , c_{\xi,2} s + c_{\xi,3},\right.\nonumber \\
 & \left.   | c_{\xi,4} s + c_{\xi,5}| , c_{\xi,6} + \theta(- \varsigma_\xi(C_\xi, s)  \pi) \right),
\end{align}
where 
\begin{equation}\label{varsigma_C_xi_s}
\varsigma_\xi(C_\xi, s)  = \text{sgn} \left( c_{\xi,4} s + c_{\xi,5} \right)
\end{equation}
indicates the specific branch the particle follows at different time intervals, before and after crossing the center at $r=0$. To establish a one-to-one correspondence between the parameters $C_\xi = (c_{\xi,1}, \dots, c_{\xi,6})$ and the geodesics, we assume $c_{\xi,6}\in[0,\pi)$.

\section{Matching conditions}\label{Matching_conditions}

In Sec.\ \ref{Local_geodesics}, we analyzed the trajectories of particles within each spacetime region ($\mathcal{R}_\text{p}$, $\mathcal{R}_\text{g}$, $\mathcal{R}_\text{a}$ and $\mathcal{R}_\text{b}$). In this section, we derive the continuity conditions for pairs of geodesics at the interfaces between neighboring regions. These conditions will enable the construction of global geodesics for particles that travel within multiple regions of the spacetime.

\subsection{Pulse-GW interface ($\mathcal{R}_\text{p} \leftrightarrow \mathcal{R}_\text{g}$)}

In Sec.\ \ref{Light_pulse}, we derived the trajectory of particles within the pulse region $\mathcal{R}_\text{p}$. Expressed in $(u, v_{\text{p}\text{g}}, r_{\text{p}\text{g}}, \varphi)$ coordinates, these geodesics are described by Eq.\ (\ref{geodesics_p_varsigma}). Later, in Sec.\ \ref{Gravitational_wave}, we studied the geodesics within the GW region $\mathcal{R}_\text{g}$, leading to Eq.\ (\ref{geodesics_g}). In this subsection, we combine these results to describe a particle moving through both regions ($\mathcal{R}_\text{p}$ and $\mathcal{R}_\text{g}$) while crossing the boundary at $r = \bar{r}_\text{p}$. By imposing the continuity conditions $\left. (u(s), v_{\text{p}\text{g}}(s), r_{\text{p}\text{g}}(s), \varphi(s)) \right|_{\mathcal{R}_\text{p}} = \left. (u(s), v_{\text{p}\text{g}}(s), r_{\text{p}\text{g}}(s), \varphi(s)) \right|_{\mathcal{R}_\text{g}}$ and $\left. U(s) \right|_{\mathcal{R}_\text{p}}= \left. U(s) \right|_{\mathcal{R}_\text{g}}$ at $r_{\text{p}\text{g}}(s) = \bar{r}_\text{p}$, we derive the one-to-one map $  (c_{\text{p},1}, \dots, c_{\text{p},6}, \sigma_\text{p}, n_\text{p}) \leftrightarrow (c_{\text{g},1}, \dots, c_{\text{g},6}) $, which allow us to identify geodesics spanning $\mathcal{R}_\text{p} \cup \mathcal{R}_\text{g}$.

First, we compute the map $(c_{\text{p},1}, \dots, c_{\text{p},6}, \sigma_\text{p}, n_\text{p}) \mapsto (c_{\text{g},1}, \dots, c_{\text{g},6})$ for geodesics in $\mathcal{R}_\text{p}$ that cross the boundary $r_\text{pg} = \bar{r}_\text{p}$. Each configuration $(c_{\text{p},1}, \dots, c_{\text{p},6}, \sigma_\text{p}, n_\text{p})$ in $\mathcal{R}_\text{p}$ has two crossing points at $r_{\text{pg}} = \bar{r}_\text{p}$ labeled by $\varsigma_\text{p} \in \{ -1, +1 \}$ [Eq.\ (\ref{s_bar_sigman})]. The extension of this geodesic into $\mathcal{R}_\text{g}$ beyond the crossing point $\varsigma_\text{p}$ is given by Eq.\ (\ref{geodesics_g}), with the parameters $(c_{\text{g},1}, \dots, c_{\text{g},6}) $ fixed by the continuity conditions $\left. (u(s), v_{\text{p}\text{g}}(s), r_{\text{p}\text{g}}(s), \varphi(s)) \right|_{\mathcal{R}_\text{p}} = \left. (u(s), v_{\text{p}\text{g}}(s), r_{\text{p}\text{g}}(s), \varphi(s)) \right|_{\mathcal{R}_\text{g}}$ and $\left. U(s) \right|_{\mathcal{R}_\text{p}}= \left. U(s) \right|_{\mathcal{R}_\text{g}}$ at $r_{\text{p}\text{g}}(s) = \bar{r}_\text{p}$. By using Eqs.\ (\ref{geodesics_p}) and (\ref{geodesics_g}) and imposing such conditions, we find
\begin{widetext}
\begin{align}\label{c_g_c_p}
& \begin{aligned} & c_{\text{g},1} =  c_{\text{p},1}, & c_{\text{g},2} =  c_{\text{p},2}, \end{aligned} \nonumber \\
& c_{\text{g},3} = c_{\text{p},3}  +  \varsigma_\text{p} \sigma_{\text{p}} \frac{\sqrt{\epsilon }}{2} \sqrt{c_{\text{p},5}^2-\bar{r}_{\text{p}}^2} -  \varsigma_\text{p} \sigma_{\text{p}} \frac{\sqrt{\pi \epsilon}}{2 \sqrt{2}} \left(2-\frac{c_{\text{p},5}^2}{\bar{r}_{\text{p}}^2}\right) \exp \left[ \frac{1}{2} \left(\frac{c_{\text{p},5}^2}{\bar{r}_{\text{p}}^2}-1\right) \right] \text{erf}\left[\sqrt{ \frac{1}{2} \left( \frac{c_{\text{p},5}^2}{\bar{r}_{\text{p}}^2}-1\right)}\right] \bar{r}_{\text{p}} , \nonumber \\
& c_{\text{g},4} = c_{\text{p},4}  -\frac{ 1 }{\sqrt{\epsilon }}  \left[ \sigma_{\text{p}} \arccos\left( \frac{ \varsigma_\text{p} \bar{r}_{\text{p}}}{ c_{\text{p},5} }\right) +2 \pi  n_{\text{p}} \right] \bar{r}_{\text{p}}  +   \varsigma_\text{p} \sigma_{\text{p}} \sqrt{\frac{\pi }{2 \epsilon}}  \exp \left[ \frac{1}{2} \left(\frac{c_{\text{p},5}^2}{\bar{r}_{\text{p}}^2}-1\right) \right] \text{erf}\left[\sqrt{ \frac{1}{2} \left( \frac{c_{\text{p},5}^2}{\bar{r}_{\text{p}}^2}-1\right)}\right] \bar{r}_{\text{p}} , \nonumber \\
& \begin{aligned} & c_{\text{g},5} =  \exp \left[ \frac{1}{2} \left(\frac{c_{\text{p},5}^2}{\bar{r}_{\text{p}}^2}-1\right) \right] \bar{r}_{\text{p}} , && c_{\text{g},6} = c_{\text{p},6} + \theta (-\varsigma_\text{p}) \pi. \end{aligned}
\end{align}

The inverse map $ (c_{\text{g},1}, \dots, c_{\text{g},6}) \mapsto (c_{\text{p},1}, \dots, c_{\text{p},6}, \sigma_\text{p}, n_\text{p}) $ is given by
\begin{align}\label{c_p_c_g}
& \begin{aligned} & c_{\text{p},1} =  c_{\text{g},1}, & c_{\text{p},2} =  c_{\text{g},2}, \end{aligned} \nonumber \\
& c_{\text{p},3} = c_{\text{g},3} -   \varsigma_\text{g} \frac{\sqrt{\pi\epsilon}}{2 \sqrt{2}}  \left[1-2 \log \left(\frac{c_{\text{g},5}}{\bar{r}_{\text{p}}}\right)\right] \text{erf}\left[\sqrt{\log \left(\frac{c_{\text{g},5}}{\bar{r}_{\text{p}}}\right)}\right]  c_{\text{g},5} + \varsigma_\text{g}  \sqrt{\frac{\epsilon}{2}  \log \left(\frac{c_{\text{g},5}}{\bar{r}_{\text{p}}}\right)} \bar{r}_{\text{p}}, \nonumber \\
& c_{\text{p},4} = c_{\text{g},4} +  \varsigma_\text{g}  \sqrt{\frac{\pi }{2 \epsilon}} \text{erf}\left[\sqrt{\log \left(\frac{c_{\text{g},5}}{\bar{r}_{\text{p}}}\right)}\right] c_{\text{g},5} + \frac{1}{\sqrt{\epsilon }} \left\lbrace- \varsigma_\text{p}(c_{\text{g},6})   \varsigma_\text{g} \arccos\left[\frac{  \varsigma_\text{p}(c_{\text{g},6})}{\sqrt{1+2 \log \left(\frac{c_{\text{g},5}}{\bar{r}_{\text{p}}}\right)}}\right] + 2 n_{\text{p}}  \pi \right\rbrace \bar{r}_{\text{p}}, \nonumber \\
& \begin{aligned} & c_{\text{p},5} =  \left[ \sqrt{ 1+ 2 \log \left(\frac{c_{\text{g},5}}{\bar{r}_{\text{p}}}\right) } \right] \bar{r}_{\text{p}} , && c_{\text{p},6} =  c_{\text{g},6} - \theta (-\varsigma_\text{p}(c_{\text{g},6})) \pi , && \sigma_{\text{p}}=- \varsigma_\text{p}(c_{\text{g},6})   \varsigma_\text{g}, \end{aligned} \nonumber \\
& n_{\text{p}} = - \left\lfloor \frac{\sqrt{\epsilon}}{2 \pi}  \frac{c_{\text{g},4}}{\bar{r}_{\text{p}}} + \frac{  \varsigma_\text{g}}{2 \sqrt{2 \pi}}    \text{erf}\left[\sqrt{\log \left(\frac{c_{\text{g},5}}{\bar{r}_{\text{p}}}\right)}\right] \frac{c_{\text{g},5}}{\bar{r}_{\text{p}}} -  \frac{ \varsigma_\text{p}(c_{\text{g},6})   \varsigma_\text{g}}{2\pi} \arccos\left[\frac{  \varsigma_\text{p}(c_{\text{g},6})}{\sqrt{1+2 \log \left(\frac{c_{\text{g},5}}{\bar{r}_{\text{p}}}\right)}}\right]  \right\rfloor.
\end{align}
\end{widetext}
In Eq.\ (\ref{c_p_c_g}), the parameter $\varsigma_\text{g}$ identifies the specific crossing point [Eq.\ (\ref{s_bar_C_g_varsigma_g})] where the continuity conditions are applied. The function $  \varsigma_\text{p}(c_{\text{g},6}) $ specifies the branch $C_{\text{p},\varsigma_\text{p}}$ within $\mathcal{R}_\text{p}$ that matches the geodesic in $\mathcal{R}_\text{g}$. If the particle crosses the boundary $r_{\text{pg}} = \bar{r}_\text{p}$ with the azimuthal coordinate $\varphi \in [0, \pi)$, then $  \varsigma_\text{p} = +1$; otherwise, if  $\varphi \in [\pi, 2\pi)$, $  \varsigma_\text{p} = -1$. Hence,
\begin{equation}
\varsigma_\text{p}(c_{\text{g},6}) = \begin{cases}
+1 & \text{if } c_{\text{g},6} \in [0, \pi),\\
-1 & \text{if } c_{\text{g},6} \in [\pi, 2\pi).
\end{cases}
\end{equation}

\subsection{Pulse-exterior interface ($\mathcal{R}_\text{p} \leftrightarrow \mathcal{R}_\xi$)}


The geodesics in $\mathcal{R}_\text{p}$, introduced in Sec.\ \ref{Light_pulse}, can be smoothly extended into the geodesics in the exterior region $\mathcal{R}_\xi$ defined by Eq.\ (\ref{geodesics_xieta}). At the boundary $u = \bar{u}_\xi$, each geodesic must satisfy the continuity conditions $\left. (u(s), v_{\text{p}\text{g}}(s), r_{\text{p}\text{g}}(s), \varphi(s)) \right|_{\mathcal{R}_\text{p}} = \left. (u(s),v_{\xi\text{p}}(s),r_{\xi\text{p}}(s),\varphi(s)) \right|_{\mathcal{R}_\xi}$ and $\left. U(s) \right|_{\mathcal{R}_\text{p}}= \left. U(s) \right|_{\mathcal{R}_\xi}$. By using these conditions, we derive the map $(c_{\text{p},1}, \dots, c_{\text{p},6}) \mapsto (c_{\xi ,1}, \dots, c_{\xi ,6})$ which specifies how any geodesic in  $\mathcal{R}_\text{p}$ [Eqs.\ (\ref{geodesics_p_varsigma})] can be extended into geodesics in $\mathcal{R}_\xi$ [Eq.\ (\ref{geodesics_xieta})]. The result is
\begin{widetext}
\begin{align}\label{c_xi_c_p}
& \begin{aligned} & c_{\xi ,1} =  c_{\text{p},1}, & c_{\xi ,2} = c_{\text{p},2} + \frac{\epsilon  c_{\text{p},1} }{4} \frac{ c_{\text{p},5}^2}{ \bar{r}_{\text{p}}^2} \left[ 1 -  \cos \left( 2 \sqrt{\epsilon } \frac{c_{\text{p},4}+ \bar{u}_{\xi }}{\bar{r}_{\text{p}}}\right) \right], \end{aligned} \nonumber \\
& c_{\xi ,3} =  c_{\text{p},3} -  \frac{\sqrt{\epsilon }}{4} \frac{ c_{\text{p},5}^2 }{ \bar{r}_{\text{p}}^2} \left\lbrace \sin \left( 2 \sqrt{\epsilon } \frac{c_{\text{p},4}+ \bar{u}_{\xi }}{\bar{r}_{\text{p}}} \right) \bar{r}_{\text{p}}  + \sqrt{\epsilon} \left[1-  \cos \left( 2 \sqrt{\epsilon } \frac{c_{\text{p},4}+ \bar{u}_{\xi }}{\bar{r}_{\text{p}}}\right)  \right] \bar{u}_{\xi } \right\rbrace, \nonumber \\
& \begin{aligned} & c_{\xi ,4} =  - \sqrt{\epsilon } c_{\text{p},1}\frac{ c_{\text{p},5}}{\bar{r}_{\text{p}}} \sin \left(\sqrt{\epsilon }\frac{c_{\text{p},4}+ \bar{u}_{\xi }}{\bar{r}_{\text{p}}}\right), & c_{\xi ,5} =\frac{ c_{\text{p},5} }{ \bar{r}_{\text{p}}} \left[ \cos \left( \sqrt{\epsilon } \frac{c_{\text{p},4}+ \bar{u}_{\xi }}{\bar{r}_{\text{p}}} \right) \bar{r}_{\text{p}} + \sqrt{\epsilon}  \sin \left( \sqrt{\epsilon } \frac{c_{\text{p},4}+ \bar{u}_{\xi }}{\bar{r}_{\text{p}}}\right)  \bar{u}_{\xi } \right],  \end{aligned} \nonumber \\
&  c_{\xi ,6} = c_{\text{p},6}.
\end{align}

The inverse transformation is given by
\begin{align}\label{c_p_c_xi}
& \begin{aligned} & c_{\text{p},1} = c_{\xi ,1}, && c_{\text{p},2} = c_{\xi ,2}-\frac{c_{\xi ,4}^2}{2 c_{\xi ,1}},  && c_{\text{p},3} = c_{\xi ,3}-\frac{c_{\xi ,4}}{2 c_{\xi ,1}} c_{\xi ,5}, \end{aligned} \nonumber \\
& c_{\text{p},4} = -\bar{u}_{\xi } +  \frac{1}{\sqrt{\epsilon }} \left\lbrace - \arctan\left( \frac{c_{\xi ,4}}{\sqrt{\epsilon } }  \frac{\bar{r}_{\text{p}}}{c_{\xi ,1} c_{\xi ,5} + c_{\xi ,4}\bar{u}_{\xi }} \right)  +2 \pi  n_\xi (C_\xi) + \theta \left[ - \left( c_{\xi ,5} + \frac{c_{\xi ,4}}{c_{\xi ,1}} \bar{u}_{\xi } \right) \right] \pi \right\rbrace \bar{r}_{\text{p}}, \nonumber \\
& \begin{aligned} & c_{\text{p},5}=  \sqrt{\frac{1}{\epsilon } \left(\frac{c_{\xi ,4}}{ c_{\xi ,1}} \bar{r}_{\text{p}} \right)^2  +   \left( c_{\xi ,5} + \frac{c_{\xi ,4}}{c_{\xi ,1}} \bar{u}_{\xi }\right)^2 }, & c_{\text{p},6} = c_{\xi ,6}, \end{aligned} 
\end{align}
with
\begin{equation}
n_{\xi}(C_\xi) = - \left\lfloor - \frac{\sqrt{\epsilon }}{2\pi} \frac{\bar{u}_{\xi }}{\bar{r}_{\text{p}}}  - \frac{1}{2\pi} \arctan\left( \frac{c_{\xi ,4}}{\sqrt{\epsilon } }  \frac{\bar{r}_{\text{p}}}{c_{\xi ,1} c_{\xi ,5} + c_{\xi ,4}\bar{u}_{\xi }} \right)  + \frac{1}{2} \theta \left[ - \left( c_{\xi ,5} + \frac{c_{\xi ,4}}{c_{\xi ,1}} \bar{u}_{\xi } \right) \right]  \right\rfloor.
\end{equation}
\end{widetext}
In the particular case where $|c_{\text{p},5}| > \bar{r}_\text{p}$, the particle crosses the boundary $r_\text{pg} = \bar{r}_\text{p}$ of $\mathcal{R}_\text{p}$ in addition to $u = \bar{u}_\xi$. Under these circumstances, it becomes necessary to express the parameters $(\sigma_\text{p}, n_\text{p})$ in terms of $(c_{\xi ,1}, \dots, c_{\xi ,6})$. To determine the mapping $ (c_{\xi ,1}, \dots, c_{\xi ,6}) \mapsto (\sigma_\text{p}, n_\text{p})$, we use the requirement that $s = \bar{u}_\xi / c_{\xi ,1}$ must lie within the domain of the geodesic in $\mathcal{R}_\text{p}$. Specifically, $\sigma_\text{p}$ and $ n_\text{p}$ are such that $ \bar{s}_{\sigma_{\text{p}},n_{\text{p}}} (C_\text{p},  \sigma_{\text{p}}) \leq \bar{u}_\xi / c_{\xi ,1} \leq \bar{s}_{\sigma_{\text{p}},n_{\text{p}}} (C_\text{p}, - \sigma_{\text{p}})$.

\subsection{GW-exterior interface ($\mathcal{R}_\text{g} \leftrightarrow \mathcal{R}_\xi$)}

In this subsection, we study the trajectory of particles crossing the boundary between the GW region $\mathcal{R}_\text{g}$ and the exterior region $\mathcal{R}_\xi$. By imposing the continuity conditions $\left. (u(s), v_{\text{p}\text{g}}(s), r_{\text{p}\text{g}}(s), \varphi(s)) \right|_{\mathcal{R}_\text{g}} = \left. (u(s),v_{\xi\text{g}}(s),r_{\xi\text{g}}(s),\varphi(s)) \right|_{\mathcal{R}_\xi}$ and $\left. U(s) \right|_{\mathcal{R}_\text{g}}= \left. U(s) \right|_{\mathcal{R}_\xi}$ at $u = \bar{u}_\xi$, we can compute the map $(c_{\text{g},1}, \dots, c_{\text{g},6}) \mapsto (c_{\xi ,1}, \dots, c_{\xi ,6})$ to extend the geodesics of $\mathcal{R}_\text{g} $ [Eq.\ (\ref{geodesics_g})] into $\mathcal{R}_\xi $.

As a result of our computation, we obtain
\begin{widetext}
\begin{align}\label{c_xi_c_g}
& \begin{aligned} & c_{\xi ,1} =  c_{\text{g},1}, & c_{\xi ,2} = c_{\text{g},2} + \epsilon  c_{\text{g},1} \left[ \text{erf}^{-1}\left( \sqrt{\frac{2 \epsilon}{\pi }}  \frac{c_{\text{g},4} + \bar{u}_{\xi }}{c_{\text{g},5}}\right) \right]^2, \end{aligned} \nonumber \\
&  \begin{aligned} c_{\xi ,3} = \,  & c_{\text{g},3} + \frac{\epsilon}{2} \left( c_{\text{g},4} + \bar{u}_{\xi } \right) -\sqrt{2 \epsilon } \exp \left\lbrace - \left[ \text{erf}^{-1}\left( \sqrt{\frac{2 \epsilon}{\pi }}  \frac{c_{\text{g},4} + \bar{u}_{\xi }}{c_{\text{g},5}}\right) \right]^2\right\rbrace \text{erf}^{-1}\left( \sqrt{\frac{2 \epsilon}{\pi }}  \frac{c_{\text{g},4} + \bar{u}_{\xi }}{c_{\text{g},5}}\right) c_{\text{g},5}  \nonumber \\
& - \epsilon \left[  \text{erf}^{-1}\left( \sqrt{\frac{2 \epsilon}{\pi }}  \frac{c_{\text{g},4} + \bar{u}_{\xi }}{c_{\text{g},5}}\right) \right]^2  \bar{u}_{\xi } - \epsilon  \log \left(\frac{c_{\text{g},5}}{\bar{r}_{\text{p}}}\right)  \left(c_{\text{g},4} + \bar{u}_{\xi }\right), \end{aligned}\nonumber \\
& c_{\xi ,4} =  - \varsigma_{\xi }(c_{\text{g},6}) \sqrt{2 \epsilon } \, c_{\text{g},1}  \text{erf}^{-1}\left( \sqrt{\frac{2 \epsilon}{\pi }}  \frac{c_{\text{g},4} + \bar{u}_{\xi }}{c_{\text{g},5}}\right), \nonumber \\
&  c_{\xi ,5} =  \varsigma_{\xi }(c_{\text{g},6}) \exp \left\lbrace - \left[ \text{erf}^{-1}\left( \sqrt{\frac{2 \epsilon}{\pi }}  \frac{c_{\text{g},4} + \bar{u}_{\xi }}{c_{\text{g},5}}\right) \right]^2\right\rbrace c_{\text{g},5}  +   \varsigma_{\xi }(c_{\text{g},6}) \sqrt{2 \epsilon } \,  \text{erf}^{-1}\left( \sqrt{\frac{2 \epsilon}{\pi }}  \frac{c_{\text{g},4} + \bar{u}_{\xi }}{c_{\text{g},5}}\right) \bar{u}_{\xi },\nonumber \\
 &  c_{\xi ,6} =   c_{\text{g},6} - \theta[- \varsigma_{\xi }(c_{\text{g},6})] \pi,
\end{align}
with
\begin{equation}
\varsigma_\xi(c_{\text{g},6}) = \begin{cases}
+1 & \text{if } c_{\text{g},6} \in [0, \pi),\\
-1 & \text{if } c_{\text{g},6} \in [\pi, 2\pi).
\end{cases}
\end{equation}
The inverse map is
\begin{align}\label{c_g_c_xi}
& \begin{aligned} & c_{\text{g},1} =  c_{\xi ,1}, & c_{\text{g},2} =  c_{\xi ,2}-\frac{c_{\xi ,4}^2}{2 c_{\xi ,1}}, \end{aligned} \nonumber \\
& \begin{aligned} c_{\text{g},3} = \, & c_{\xi ,3} -\frac{c_{\xi ,4}}{c_{\xi ,1}}  \left(   c_{\xi ,5} + \frac{ c_{\xi ,4}}{2 c_{\xi ,1}} \bar{u}_{\xi } \right) \nonumber \\
&   -  \frac{\sqrt{\pi}}{2 \sqrt{2 \epsilon }}\exp \left( \frac{c_{\xi ,4}^2}{2 \epsilon  c_{\xi ,1}^2} \right) \text{erf}\left(\frac{c_{\xi ,4}}{\sqrt{2 \epsilon } c_{\xi ,1}}\right) \left[\frac{c_{\xi ,4}^2}{ c_{\xi ,1}^2}-\epsilon + 2 \epsilon \log \left( \left| c_{\xi ,5} + \frac{ c_{\xi ,4}}{c_{\xi ,1}  } \bar{u}_{\xi }\right| \frac{1}{\bar{r}_\text{p}} \right) \right]\left( c_{\xi ,5} + \frac{ c_{\xi ,4}}{c_{\xi ,1}  } \bar{u}_{\xi } \right),  \end{aligned} \nonumber \\
& \begin{aligned}  & c_{\text{g},4} = -\bar{u}_{\xi }  - \sqrt{\frac{\pi }{2\epsilon}} \exp \left( \frac{c_{\xi ,4}^2}{2 \epsilon  c_{\xi ,1}^2} \right)   \text{erf}\left(\frac{c_{\xi ,4}}{\sqrt{2\epsilon } c_{\xi ,1}}\right) \left(   c_{\xi ,5} + \frac{ c_{\xi ,4}}{ c_{\xi ,1}} \bar{u}_{\xi } \right), & c_{\text{g},5} =  \exp \left( \frac{c_{\xi ,4}^2}{2 \epsilon  c_{\xi ,1}^2} \right)  \left|  c_{\xi ,5} + \frac{ c_{\xi ,4}}{ c_{\xi ,1}} \bar{u}_{\xi } \right| , \end{aligned} \nonumber \\
 &  c_{\text{g} ,6} =   c_{\xi,6} + \theta \left(-c_{\xi ,5}-\frac{c_{\xi ,4}}{c_{\xi ,1}} \bar{u}_{\xi } \right) \pi.
\end{align}
\end{widetext}

In the specific case where the particle is restricted to $\varphi = 0$, corresponding to configurations with $c_{\text{g} ,6} = c_{\xi ,6} = 0$, the results match those presented in Ref.\ \cite{ylvn-3ybm}.

\section{Global geodesics}\label{Geodesics}

In Sec.\ \ref{Local_geodesics}, we described the geodesics within each individual region ($\mathcal{R}_\text{p}$, $\mathcal{R}_\text{g}$, $\mathcal{R}_\text{a}$ and $\mathcal{R}_\text{b}$). Then, in Sec.\ \ref{Matching_conditions}, we derived the matching conditions at each boundary to describe the trajectories of particles transitioning between neighboring regions. In this section, we consider the spacetime as a whole and combine the results from the previous sections to study the trajectories of particles traveling through multiple regions. We focus on two specific scenarios. In the first one, the particle is directly hit by the pulse and interacts with the gravitational field confined to $\mathcal{R}_\text{p}$. In this scenario, the trajectory sequentially passes through the regions $\mathcal{R}_\text{a}$, $\mathcal{R}_\text{p}$ and $\mathcal{R}_\text{b}$. In the second scenario, the particle is scattered by the GW surrounding the pulse, resulting in a trajectory that traverses the regions $\mathcal{R}_\text{a}$, $\mathcal{R}_\text{g}$ and $\mathcal{R}_\text{b}$. 

\subsection{Particle-pulse interaction ($\mathcal{R}_\text{a} \to \mathcal{R}_\text{p} \to \mathcal{R}_\text{b}$)}\label{Particle_pulse_interaction}

In this subsection, we study the trajectory of a particle perturbed by the interaction with the gravitational field in $\mathcal{R}_\text{p}$. The particle starts its motion in the Minkowski region $\mathcal{R}_\text{a}$, moves into the pulse region $\mathcal{R}_\text{p}$, and then exits $\mathcal{R}_\text{p}$ to enter the other Minkowski region $\mathcal{R}_\text{b}$. During its entire trajectory, the particle does not enter the GW region $\mathcal{R}_\text{g}$.

Within each region ($ \mathcal{R}_\text{p}$ and $ \mathcal{R}_\xi$), the geodesics are locally described by the respective equations (\ref{geodesics_p_varsigma}) and (\ref{geodesics_xieta}), with the parameters $(c_{\text{p},1}, \dots, c_{\text{p},6})$ and $(c_{\xi,1}, \dots, c_{\xi,6})$ constrained by the matching conditions (\ref{c_xi_c_p}) and (\ref{c_p_c_xi}). The particular geodesic in $ \mathcal{R}_\text{a} \cup \mathcal{R}_\text{p} \cup \mathcal{R}_\text{b}$ describing the trajectory of the particle is identified by some fixed initial conditions. For instance, we can assume that at moment in which the particle crosses the boundary between $\mathcal{R}_\text{a}$ and $\mathcal{R}_\text{p}$ (i.e., at $u=\bar{u}_\text{a}$), its Minkowski coordinates are $(u,v,r,\varphi) = (\bar{u}_\text{a},\bar{v}_\text{a},\bar{r}_\text{a},\bar{\varphi}_\text{a})$ and its four-velocity is $U = U_{\text{a}}^u \partial_u + U_{\text{a}}^v \partial_v + U_{\text{a}}^r \partial_r$. In this case, the parameters $(c_{\text{a},1}, \dots, c_{\text{a},6})$ are given by
\begin{align}\label{c_a_initial}
& \begin{aligned} & c_{\text{a},1} = U_{\text{a}}^u, && c_{\text{a},2} = U_{\text{a}}^v, && c_{\text{a},3} = \bar{v}_{\text{a}}-\frac{U_{\text{a}}^v}{U_{\text{a}}^u} \bar{u}_{\text{a}}, \end{aligned} \nonumber \\
& \begin{aligned} & c_{\text{a},4} = \varsigma_{\text{a}} (\bar{\varphi }_{\text{a}}) U_{\text{a}}^r, && c_{\text{a},5} = \varsigma_{\text{a}} (\bar{\varphi }_{\text{a}}) \left( \bar{r}_{\text{a}}-\frac{U_{\text{a}}^r}{U_{\text{a}}^u} \bar{u}_{\text{a}} \right), \end{aligned} \nonumber \\
&  c_{\text{a},6} = \bar{\varphi}_{\text{a}} - \theta[-\varsigma_{\text{a}} (\bar{\varphi }_{\text{a}})] \pi .
\end{align}

At the moment in which the particle exits the pulse, its coordinates are labeled as $(u,v,r,\varphi) = (\bar{u}_\text{b},\bar{v}_\text{b},\bar{r}_\text{b},\bar{\varphi}_\text{b})$ and its four-velocity as $U = U_{\text{b}}^u \partial_u + U_{\text{b}}^v \partial_v + U_{\text{b}}^r \partial_r$. By using Eq.\ (\ref{geodesics_xieta}) with $\xi = \text{b}$, we obtain the map $ (c_{\text{b},1}, \dots, c_{\text{b},6}) \mapsto (\bar{v}_\text{b},\bar{r}_\text{b},\bar{\varphi}_\text{b},U_{\text{b}}^u, U_{\text{b}}^v , U_{\text{b}}^r)$, which reads
\begin{align}\label{final_c_b}
& \begin{aligned} & \bar{v}_\text{b} = c_{\text{b},3} + \frac{c_{\text{b},2}}{c_{\text{b},1}} \bar{u}_{\text{b}} , && \bar{r}_\text{b} = \left| c_{\text{b},5} + \frac{ c_{\text{b},4}}{c_{\text{b},1}} \bar{u}_{\text{b}}\right|,  \end{aligned} \nonumber \\
& \begin{aligned} & \bar{\varphi}_\text{b} = c_{\text{b},6} +  \theta \left(-c_{\text{b},5} - \frac{ c_{\text{b},4}}{c_{\text{b},1}} \bar{u}_{\text{b}}\right)  \pi, && U_{\text{b}}^u = c_{\text{b},1},  \end{aligned} \nonumber \\
& \begin{aligned} & U_{\text{b}}^v = c_{\text{b},2}, && U_{\text{b}}^r = \text{sgn}\left(c_{\text{b},5} + \frac{ c_{\text{b},4}}{c_{\text{b},1}} \bar{u}_{\text{b}}\right) c_{\text{b},4}.  \end{aligned}
\end{align}

By plugging Eq.\ (\ref{c_a_initial}) into Eq.\ (\ref{c_p_c_xi}) for $\xi=\text{a}$ and substituting the resulting equation into Eq.\ (\ref{c_xi_c_p}) for $\xi=\text{b}$ we can express the parameters $(c_{\text{b},1}, \dots, c_{\text{b},6})$  in terms of the initial conditions $(\bar{v}_\text{a},\bar{r}_\text{a},\bar{\varphi}_\text{a},U_{\text{a}}^u, U_{\text{a}}^v , U_{\text{a}}^r)$. This result can be used in Eq.\ (\ref{final_c_b}) to compute the mapping $(\bar{v}_\text{a},\bar{r}_\text{a},\bar{\varphi}_\text{a},U_{\text{a}}^u, U_{\text{a}}^v , U_{\text{a}}^r) \mapsto (\bar{v}_\text{b},\bar{r}_\text{b},\bar{\varphi}_\text{b},U_{\text{b}}^u, U_{\text{b}}^v , U_{\text{b}}^r)$ as
\begin{widetext}
\begin{align}\label{final_pulse_initial}
&  \begin{aligned} \bar{v}_\text{b} = \, & \bar{v}_{\text{a}}  + \left[ \frac{U_{\text{a}}^v}{U_{\text{a}}^u} - \frac{1}{2} \left(\frac{U_{\text{a}}^r}{U_{\text{a}}^u}\right)^2  \right] \left( \bar{u}_{\text{b}} - \bar{u}_{\text{a}} \right) - \frac{1}{2} \frac{U_{\text{a}}^r}{U_{\text{a}}^u}  \bar{r}_{\text{a}} \nonumber \\
&  + \frac{1}{4 \sqrt{\epsilon } } \sin \left[-2 \sqrt{\epsilon }  \frac{\bar{u}_{\text{b}}-\bar{u}_{\text{a}}}{\bar{r}_{\text{p}}}+ 2 \arctan \left(\frac{1}{\sqrt{\epsilon}} \frac{\bar{r}_{\text{p}}}{\bar{r}_{\text{a}}} \frac{U_{\text{a}}^r}{U_{\text{a}}^u} \right)\right] \left[\left(\frac{U_{\text{a}}^r}{U_{\text{a}}^u}\right)^2 \bar{r}_{\text{p}} + \epsilon  \frac{\bar{r}_{\text{a}}^2}{\bar{r}_{\text{p}}} \right], \end{aligned} \nonumber \\
&  \bar{r}_\text{b} =  \left| \cos \left[- \sqrt{\epsilon }  \frac{\bar{u}_{\text{b}}-\bar{u}_{\text{a}}}{\bar{r}_{\text{p}}} + \arctan \left(\frac{1}{\sqrt{\epsilon}} \frac{\bar{r}_{\text{p}}}{\bar{r}_{\text{a}}} \frac{U_{\text{a}}^r}{U_{\text{a}}^u} \right)\right]\right| \sqrt{ \frac{1}{\epsilon} \left( \frac{U_{\text{a}}^r}{U_{\text{a}}^u} \bar{r}_{\text{p}} \right)^2 +\bar{r}_{\text{a}}^2}, \nonumber \\
&  \bar{\varphi}_\text{b} = \bar{\varphi }_{\text{a}} - \theta [-\varsigma_{\text{a}}(\bar{\varphi }_{\text{a}})] \pi +   \theta \left\lbrace - \varsigma_{\text{a}} (\bar{\varphi }_{\text{a}} ) \cos \left[- \sqrt{\epsilon }  \frac{\bar{u}_{\text{b}}-\bar{u}_{\text{a}}}{\bar{r}_{\text{p}}} + \arctan \left(\frac{1}{\sqrt{\epsilon}} \frac{\bar{r}_{\text{p}}}{\bar{r}_{\text{a}}} \frac{U_{\text{a}}^r}{U_{\text{a}}^u} \right)\right] \right\rbrace \pi, \nonumber \\
& \begin{aligned}  & U_{\text{b}}^u = U_{\text{a}}^u, &  U_{\text{b}}^v = U_{\text{a}}^v -  \frac{\left(U_{\text{a}}^r\right)^2}{4 U_{\text{a}}^u}  + \frac{\epsilon}{4}  \frac{\bar{r}_{\text{a}}^2}{ \bar{r}_{\text{p}}^2} U_{\text{a}}^u   -\frac{1}{4 }\left[ \frac{\left(U_{\text{a}}^r\right)^2}{U_{\text{a}}^u} + \epsilon  \frac{\bar{r}_{\text{a}}^2}{\bar{r}_{\text{p}}^2} U_{\text{a}}^u \right] \cos \left[- 2 \sqrt{\epsilon }  \frac{\bar{u}_{\text{b}}-\bar{u}_{\text{a}}}{\bar{r}_{\text{p}}}+ 2 \arctan \left(\frac{1}{\sqrt{\epsilon}} \frac{\bar{r}_{\text{p}}}{\bar{r}_{\text{a}}} \frac{U_{\text{a}}^r}{U_{\text{a}}^u} \right)\right], \end{aligned} 
 \nonumber \\
&  U_{\text{b}}^r = \text{sgn}\left\lbrace \cos \left[- \sqrt{\epsilon }  \frac{\bar{u}_{\text{b}}-\bar{u}_{\text{a}}}{\bar{r}_{\text{p}}} + \arctan \left(\frac{1}{\sqrt{\epsilon}} \frac{\bar{r}_{\text{p}}}{\bar{r}_{\text{a}}} \frac{U_{\text{a}}^r}{U_{\text{a}}^u} \right)\right]\right\rbrace  \sin \left[- \sqrt{\epsilon }  \frac{\bar{u}_{\text{b}}-\bar{u}_{\text{a}}}{\bar{r}_{\text{p}}}+\arctan \left(\frac{1}{\sqrt{\epsilon}} \frac{\bar{r}_{\text{p}}}{\bar{r}_{\text{a}}} \frac{U_{\text{a}}^r}{U_{\text{a}}^u} \right)\right] \sqrt{\frac{ \left(U_{\text{a}}^r\right){}^2}{\left(U_{\text{a}}^u\right){}^2}+\epsilon  \frac{\bar{r}_{\text{a}}^2}{\bar{r}_{\text{p}}^2}} U_{\text{a}}^u. 
\end{align}
Such an equations shows how the trajectory of the particle is perturbed by the passage of the pulse.

In the limit $\epsilon \to 0$, Eq.\ (\ref{final_pulse_initial}) simplifies to 
\begin{align}\label{final_pulse_initial_epsilon}
&  \bar{v}_\text{b} = \bar{v}_{\text{a}} +  \frac{U_{\text{a}}^v}{U_{\text{a}}^u} \left(\bar{u}_{\text{b}}-\bar{u}_{\text{a}}\right) - \epsilon   \left[ \frac{\left(\bar{u}_{\text{b}}-\bar{u}_{\text{a}}\right) \bar{r}_{\text{a}}}{ \bar{r}_{\text{p}}^2 } \frac{U_{\text{a}}^r}{U_{\text{a}}^u}  + \frac{1}{3} \left(\frac{\bar{u}_{\text{b}}-\bar{u}_{\text{a}}}{ \bar{r}_{\text{p}} }\right)^2 \left(\frac{U_{\text{a}}^r}{U_{\text{a}}^u}\right)^2+ \frac{1}{2} \frac{\bar{r}_{\text{a}}^2}{ \bar{r}_{\text{p}}^2 } \right]    \left(\bar{u}_{\text{b}}-\bar{u}_{\text{a}}\right)  + O \left( \epsilon^2 \right), \nonumber  \\
&  \bar{r}_\text{b} =  \left|\bar{r}_{\text{a}} + \frac{U_{\text{a}}^r}{  U_{\text{a}}^u } \left( \bar{u}_{\text{b}} - \bar{u}_{\text{a}}\right) \right|  - \frac{\epsilon}{2} \text{sgn}\left[\bar{r}_{\text{a}} + \frac{U_{\text{a}}^r}{  U_{\text{a}}^u } \left( \bar{u}_{\text{b}} - \bar{u}_{\text{a}}\right) \right] \left(\frac{\bar{u}_{\text{b}}-\bar{u}_{\text{a}}}{ \bar{r}_{\text{p}}}\right)^2 \left[ \bar{r}_{\text{a}} + \frac{U_{\text{a}}^r}{ 3 U_{\text{a}}^u } \left( \bar{u}_{\text{b}} - \bar{u}_{\text{a}}\right)  \right] + O \left( \epsilon^2 \right) , \nonumber \\
&  \bar{\varphi}_\text{b} = \bar{\varphi }_{\text{a}}  - \theta [-\varsigma _{\text{a}}(\bar{\varphi }_{\text{a}})] \pi + \theta \left\lbrace -\varsigma _{\text{a}}(\bar{\varphi }_{\text{a}}) \left[\bar{r}_{\text{a}} + \frac{U_{\text{a}}^r}{  U_{\text{a}}^u } \left( \bar{u}_{\text{b}} - \bar{u}_{\text{a}}\right) \right] \right\rbrace \pi, \nonumber \\
& \begin{aligned}  & U_{\text{b}}^u = U_{\text{a}}^u, &  U_{\text{b}}^v = U_{\text{a}}^v - \epsilon  \frac{\bar{u}_{\text{b}}-\bar{u}_{\text{a}}}{ \bar{r}_{\text{p}}^2} \left[\bar{r}_{\text{a}} + \frac{U_{\text{a}}^r}{2 U_{\text{a}}^u} \left( \bar{u}_{\text{b}} -\bar{u}_{\text{a}} \right)  \right] U_{\text{a}}^r  + O \left( \epsilon^2 \right), \end{aligned} 
 \nonumber \\
&  U_{\text{b}}^r = \text{sgn}\left[\bar{r}_{\text{a}} + \frac{U_{\text{a}}^r}{U_{\text{a}}^u} \left( \bar{u}_{\text{b}} -\bar{u}_{\text{a}} \right) \right] \left\lbrace U_{\text{a}}^r- \epsilon \frac{  \bar{u}_{\text{b}}-\bar{u}_{\text{a}}}{\bar{r}_{\text{p}}^2} \left[ \bar{r}_{\text{a}} + \frac{U_{\text{a}}^r}{2 U_{\text{a}}^u} \left( \bar{u}_{\text{b}}-\bar{u}_{\text{a}} \right) \right] U_{\text{a}}^u \right\rbrace  + O \left( \epsilon^2 \right) . 
\end{align}
\end{widetext}
In the specific case where $U_{\text{a}}^r = 0$, the perturbations in $U_{\text{b}}^v$ are of the order $\epsilon^2$, whereas the perturbations in the radial velocity scale as $\epsilon$, making the radial velocity the dominant perturbative effect.

As a example, we consider a gamma ray burst, modeled as a cylinder of uniform energy density, with the following parameters: $I = P / \pi  \bar{r}_{\text{p}}^2$, $\bar{u}_{\text{b}} - \bar{u}_{\text{a}} = c \tau / \sqrt{2}$, $\bar{r}_{\text{p}} = c \tau / 100$, $P = 10^{45} \text{ W}$, $\tau = 1 \text{ s}$. Here, $P$ is the power of the pulse and $\tau$ its duration. The corresponding value of $\epsilon$ is very small ($\epsilon \sim 10^{-7}$), which justifies the approximation $\epsilon \ll 1$. We consider a test particle that is initially at rest, with $U_{\text{a}}^u = U_{\text{a}}^v = 1/\sqrt{2}$ and $U_{\text{a}}^r = 0$. When the particle is hit by the pulse, it interacts with the local gravitational field, acquiring nonzero radial velocity. This velocity is of the order of
\begin{equation}
\frac{d r}{d t}= \frac{\sqrt{2} U_{\text{b}}^r c}{U_{\text{b}}^v + U_{\text{b}}^u} \sim - 10^{-5} \frac{\bar{r}_{\text{a}}}{\bar{r}_{\text{p}}} c.
\end{equation}
Additionally, the particle experiences a radial displacement of the order of
\begin{equation}
\bar{r}_{\text{b}} - \bar{r}_{\text{a}} \sim - \frac{\bar{r}_{\text{a}}}{\bar{r}_{\text{p}}} 10^3 \text{m}
\end{equation}

\subsection{Particle-GW interaction  ($\mathcal{R}_\text{a} \to \mathcal{R}_\text{g} \to \mathcal{R}_\text{b}$)}\label{Particle_GW_interaction}

In this subsection, we describe the trajectory of a particle that is hit by the GW while never crossing the boundary between $\mathcal{R}_\text{p}$ and $\mathcal{R}_\text{g}$, thus traveling through $\mathcal{R}_\text{a}$, $ \mathcal{R}_\text{g}$ and $ \mathcal{R}_\text{b}$. The GW encounters the particle at the event $(u, v, r, \varphi) = (\bar{u}_{\text{a}}, \bar{v}_{\text{a}}, \bar{r}_{\text{a}}, \bar{\varphi}_\text{a})$, with an initial four-velocity given by $U = U_{\text{a}}^u \partial_u + U_{\text{a}}^v \partial_v + U_{\text{a}}^r \partial_r$. After the passage of the GW, the particle reaches the point $(u,v,r,\varphi) = (\bar{u}_\text{b},\bar{v}_\text{b},\bar{r}_\text{b},\bar{\varphi}_\text{b})$ with a new four-velocity $U = U_{\text{b}}^u \partial_u + U_{\text{b}}^v \partial_v + U_{\text{b}}^r \partial_r$. The mapping $(\bar{v}_\text{a},\bar{r}_\text{a},\bar{\varphi}_\text{a},U_{\text{a}}^u, U_{\text{a}}^v , U_{\text{a}}^r) \mapsto (\bar{v}_\text{b},\bar{r}_\text{b},\bar{\varphi}_\text{b},U_{\text{b}}^u, U_{\text{b}}^v , U_{\text{b}}^r)$ can be computed similarly to Sec.\ \ref{Particle_pulse_interaction}, but with Eqs.\ (\ref{c_xi_c_g}) and (\ref{c_g_c_xi}) instead of Eqs.\ (\ref{c_xi_c_p}) and (\ref{c_p_c_xi}), resulting in
\begin{widetext}
\begin{align}\label{final_GW_initial}
& \begin{aligned} \bar{v}_\text{b} = \, & \bar{v}_{\text{a}}  - \frac{U_{\text{a}}^r}{U_{\text{a}}^u} \bar{r}_{\text{a}}  + \left[ \frac{U_{\text{a}}^v}{U_{\text{a}}^u} - \left(\frac{U_{\text{a}}^r}{U_{\text{a}}^u}\right)^2 +  \frac{\epsilon}{2} -  \epsilon \log \left(\frac{\bar{r}_{\text{a}}}{\bar{r}_{\text{p}}} \right) \right] \left( \bar{u}_{\text{b}} -  \bar{u}_{\text{a}} \right)  + \sqrt{2 \epsilon } \, \text{erf}^{-1}\left\lbrace - \sqrt{\frac{2 \epsilon}{\pi}}  \exp\left[ - \frac{1}{2 \epsilon}  \left( \frac{U_{\text{a}}^r}{ U_{\text{a}}^u} \right)^2  \right]   \frac{\bar{u}_{\text{b}}-\bar{u}_{\text{a}}}{\bar{r}_{\text{a}}} \right. \nonumber \\
 & \left. + \text{erf}\left( \sqrt{\frac{1}{2 \epsilon} }  \frac{U_{\text{a}}^r}{U_{\text{a}}^u}\right)\right\rbrace  \exp \left[ \frac{1}{2 \epsilon  } \left( \frac{U_{\text{a}}^r}{U_{\text{a}}^u} \right)^2 - \left( \text{erf}^{-1}\left\lbrace - \sqrt{\frac{2 \epsilon}{\pi}}  \exp\left[ - \frac{1}{2 \epsilon}  \left( \frac{U_{\text{a}}^r}{ U_{\text{a}}^u} \right)^2  \right]   \frac{\bar{u}_{\text{b}}-\bar{u}_{\text{a}}}{\bar{r}_{\text{a}}} + \text{erf}\left( \sqrt{\frac{1}{2 \epsilon} }  \frac{U_{\text{a}}^r}{U_{\text{a}}^u}\right)\right\rbrace \right)^2 \right] \bar{r}_{\text{a}}, \end{aligned}  \nonumber \\
& \begin{aligned} & \bar{r}_\text{b} =  \exp \left[ \frac{1}{2 \epsilon  } \left( \frac{U_{\text{a}}^r}{U_{\text{a}}^u} \right)^2 - \left( \text{erf}^{-1}\left\lbrace - \sqrt{\frac{2 \epsilon}{\pi}}  \exp\left[ - \frac{1}{2 \epsilon}  \left( \frac{U_{\text{a}}^r}{ U_{\text{a}}^u} \right)^2  \right]   \frac{\bar{u}_{\text{b}}-\bar{u}_{\text{a}}}{\bar{r}_{\text{a}}} + \text{erf}\left( \sqrt{\frac{1}{2 \epsilon} }  \frac{U_{\text{a}}^r}{U_{\text{a}}^u}\right)\right\rbrace \right)^2 \right] \bar{r}_{\text{a}}, && \bar{\varphi }_{\text{b}} =  \bar{\varphi }_{\text{a}}, \end{aligned}  \nonumber \\
& \begin{aligned} & U_{\text{b}}^u = U_{\text{a}}^u, && U_{\text{b}}^v = U_{\text{a}}^v -\frac{\left(U_{\text{a}}^r\right)^2}{2 U_{\text{a}}^u}+ \epsilon  \left( \text{erf}^{-1}\left\lbrace - \sqrt{\frac{2 \epsilon}{\pi}}  \exp\left[ - \frac{1}{2 \epsilon}  \left( \frac{U_{\text{a}}^r}{ U_{\text{a}}^u} \right)^2  \right]   \frac{\bar{u}_{\text{b}}-\bar{u}_{\text{a}}}{\bar{r}_{\text{a}}} + \text{erf}\left( \sqrt{\frac{1}{2 \epsilon} }  \frac{U_{\text{a}}^r}{U_{\text{a}}^u}\right)\right\rbrace \right)^2 U_{\text{a}}^u ,  \end{aligned} \nonumber \\
& U_{\text{b}}^r = \sqrt{2 \epsilon }  \text{erf}^{-1}\left\lbrace - \sqrt{\frac{2 \epsilon}{\pi}}  \exp\left[ - \frac{1}{2 \epsilon}  \left( \frac{U_{\text{a}}^r}{ U_{\text{a}}^u} \right)^2  \right]   \frac{\bar{u}_{\text{b}}-\bar{u}_{\text{a}}}{\bar{r}_{\text{a}}} + \text{erf}\left( \sqrt{\frac{1}{2 \epsilon} }  \frac{U_{\text{a}}^r}{U_{\text{a}}^u}\right)\right\rbrace U_{\text{a}}^u.
\end{align}
\end{widetext}

To examine the limit $\epsilon \to 0$ in Eq.\ (\ref{final_GW_initial}) we consider three distinct cases: $U_{\text{a}}^r = 0$, $U_{\text{a}}^r \sim \epsilon$ and $U_{\text{a}}^r \sim 1$. In the case where $U_{\text{a}}^r = 0$,  we obtain
\begin{align}
&   \begin{aligned} \bar{v}_\text{b} = \, & \bar{v}_{\text{a}} + \frac{U_{\text{a}}^v}{U_{\text{a}}^u} \left( \bar{u}_{\text{b}} - \bar{u}_{\text{a}} \right)  - \epsilon \left[ \frac{1}{2} + \log \left( \frac{\bar{r}_{\text{a}}}{\bar{r}_{\text{p}}}\right) \right] \left( \bar{u}_{\text{b}} - \bar{u}_{\text{a}} \right) \nonumber \\
&  + O\left(\epsilon^2\right),\end{aligned} \nonumber \\
& \begin{aligned} & \bar{r}_\text{b} = \bar{r}_{\text{a}}  - \frac{\epsilon}{2} \frac{\left( \bar{u}_{\text{b}} - \bar{u}_{\text{a}} \right) ^2}{\bar{r}_{\text{a}}} + O\left(\epsilon^2\right),  && \bar{\varphi }_{\text{b}} =  \bar{\varphi }_{\text{a}}, \end{aligned} \nonumber \\
& \begin{aligned} & U_{\text{b}}^u = U_{\text{a}}^u, && U_\text{b}^v = U_{\text{a}}^v +\frac{\epsilon^2}{2} \left(\frac{ \bar{u}_{\text{b}}-\bar{u}_{\text{a}}}{\bar{r}_{\text{a}}}\right)^2 U_{\text{a}}^u + O\left(\epsilon^3\right),  \end{aligned} \nonumber \\
& U_\text{b}^r = -\epsilon  \frac{ \bar{u}_{\text{b}} - \bar{u}_{\text{a}}}{\bar{r}_{\text{a}}} U_{\text{a}}^u + O\left(\epsilon^2\right).
\end{align}
In the specific case where $ \bar{\varphi }_{\text{a}} = 0$ and $U_{\text{a}}^u=U_{\text{a}}^v=1/\sqrt{2}$, this result matches the one presented in Ref.\ \cite{ylvn-3ybm}.

If $U_{\text{a}}^r$ scales as $U_{\text{a}}^r \sim \epsilon$ in the limit $\epsilon \to 0$, we find
\begin{align}
&   \begin{aligned} \bar{v}_\text{b} = \, & \bar{v}_{\text{a}} + \frac{U_{\text{a}}^v}{U_{\text{a}}^u} \left( \bar{u}_{\text{b}} - \bar{u}_{\text{a}} \right)  - \epsilon \left[ \frac{1}{2} + \log \left( \frac{\bar{r}_{\text{a}}}{\bar{r}_{\text{p}}}\right) \right] \left( \bar{u}_{\text{b}} - \bar{u}_{\text{a}} \right) \nonumber \\
&  + O\left(\epsilon^2\right),\end{aligned} \nonumber \\
& \begin{aligned} & \bar{r}_\text{b} = \bar{r}_{\text{a}} +  \frac{U_{\text{a}}^r }{U_{\text{a}}^u}\left( \bar{u}_{\text{b}} - \bar{u}_{\text{a}} \right)  - \frac{\epsilon}{2} \frac{\left( \bar{u}_{\text{b}} - \bar{u}_{\text{a}} \right) ^2}{\bar{r}_{\text{a}}} + O\left(\epsilon^2\right),  \end{aligned}\nonumber \\
& \begin{aligned} & \bar{\varphi }_{\text{b}} =  \bar{\varphi }_{\text{a}}, && U_{\text{b}}^u = U_{\text{a}}^u,\end{aligned} \nonumber \\
& U_\text{b}^v = U_{\text{a}}^v- \epsilon  \frac{ \bar{u}_{\text{b}} - \bar{u}_{\text{a}}}{\bar{r}_{\text{a}}} U_{\text{a}}^r +\frac{\epsilon^2}{2} \left(\frac{ \bar{u}_{\text{b}}-\bar{u}_{\text{a}}}{\bar{r}_{\text{a}}}\right)^2 U_{\text{a}}^u + O\left(\epsilon^3\right), \nonumber \\  
& U_\text{b}^r = U_{\text{a}}^r-\epsilon  \frac{ \bar{u}_{\text{b}} - \bar{u}_{\text{a}}}{\bar{r}_{\text{a}}} U_{\text{a}}^u + O\left(\epsilon^2\right).
\end{align}

The case $U_{\text{a}}^r \sim 1$ (i.e., when $U_{\text{a}}^r $ is nonzero and remains unaffected as $\epsilon \to 0$) is not straightforward. To evaluate the limit of Eq.\ (\ref{final_GW_initial}), we use the expansion of the inverse complementary error function $\text{erfc}^{-1}(x_\epsilon) = \text{erf}^{-1}(1-x_{\epsilon })$ in the limit $x_\epsilon \to 0^+$ \cite{NIST:DLMF}, given by
\begin{equation}
\text{erfc}^{-1}(x_\epsilon) \approx u_{\epsilon }^{-1/2} + a_{\epsilon ,2} u_{\epsilon }^{3/2} + a_{\epsilon ,3} u_{\epsilon }^{5/2} + \dots,
\end{equation}
where
\begin{align}
& \begin{aligned} & u_{\epsilon }=-\frac{2}{\log [-\pi  x_{\epsilon }^2 \log ( x_{\epsilon })]}, &&  v_{\epsilon }=\log [-\pi \log (x_{\epsilon })]-2,\end{aligned} \nonumber \\
& \begin{aligned}  & a_{\epsilon ,2}=\frac{v_{\epsilon }}{8}, && a_{\epsilon ,3}=-\frac{1}{32}(v_{\epsilon }^2+6 v_{\epsilon }-6). \end{aligned}
\end{align}
By setting
\begin{align}
x_\epsilon = & \exp \left[ -\frac{1}{2 \epsilon} \left(\frac{U_{\text{a}}^r}{U_{\text{a}}^u} \right)^2 \right] \left[ \text{sgn}\left(\frac{U_{\text{a}}^u}{U_{\text{a}}^r} \right) \sqrt{\frac{2 \epsilon}{\pi }}  \left( \frac{ \bar{u}_{\text{b}} -\bar{u}_{\text{a}}}{\bar{r}_{\text{a}}} \right. \right. \nonumber \\
& \left. \left. +  \frac{U_{\text{a}}^u}{ U_{\text{a}}^r} \right) + O\left( \epsilon^{3/2} \right) \right],
\end{align}
we derive the expansion
\begin{widetext}
\begin{align}\label{erf_expansion}
& \text{erf}^{-1}\left\lbrace \sqrt{\frac{2 \epsilon}{\pi}}  \exp\left[ - \frac{1}{2 \epsilon}  \left( \frac{U_{\text{a}}^r}{ U_{\text{a}}^u} \right)^2  \right]   \frac{\bar{u}_{\text{b}}-\bar{u}_{\text{a}}}{\bar{r}_{\text{a}}} - \text{erf}\left( \sqrt{\frac{1}{2 \epsilon} }  \frac{U_{\text{a}}^r}{U_{\text{a}}^u}\right)\right\rbrace = -\frac{1}{\sqrt{2 \epsilon } }\frac{U_{\text{a}}^r}{U_{\text{a}}^u} + \sqrt{\frac{\epsilon }{2} } \log \left(1 + \frac{\bar{u}_{\text{b}} -\bar{u}_{\text{a}} }{\bar{r}_{\text{a}} } \frac{ U_{\text{a}}^r}{ U_{\text{a}}^u} \right)  \frac{U_{\text{a}}^u}{U_{\text{a}}^r}  \nonumber \\
&  + \frac{\epsilon^{3/2}}{\sqrt{2}}  \left\lbrace 1  - \frac{1}{2} \log \left[\frac{\pi }{2 \epsilon } \left(\frac{U_{\text{a}}^r}{U_{\text{a}}^u}\right)^2 \right]  + \frac{1}{2}\left[ \log\left( 1 + \frac{\bar{u}_{\text{b}} -\bar{u}_{\text{a}} }{\bar{r}_{\text{a}} } \frac{ U_{\text{a}}^r}{ U_{\text{a}}^u}  \right) \right]^2 \right\rbrace\left(\frac{U_{\text{a}}^u}{U_{\text{a}}^r}\right)^3 +  O\left(\epsilon^{5/2}\right).
\end{align}
\end{widetext}

Equation (\ref{erf_expansion}) can be used in Eq.\ (\ref{final_GW_initial}) to obtain
\begin{align}
& \begin{aligned} \bar{v}_\text{b} = \, & \bar{v}_{\text{a}} + \frac{U_{\text{a}}^v}{U_{\text{a}}^u} \left( \bar{u}_{\text{b}} - \bar{u}_{\text{a}} \right)\nonumber \\
& + \frac{\epsilon  \log (\epsilon ) }{2} \left(\bar{u}_{\text{b}} - \bar{u}_{\text{a}} + \frac{U_{\text{a}}^u}{U_{\text{a}}^r} \bar{r}_{\text{a}} \right) + O(\epsilon) , \end{aligned} \nonumber \\
& \begin{aligned} \bar{r}_\text{b} = \, & \bar{r}_{\text{a}} +  \frac{U_{\text{a}}^r }{U_{\text{a}}^u} \left( \bar{u}_{\text{b}} - \bar{u}_{\text{a}} \right)   \nonumber \\
&  +  \frac{\epsilon  \log (\epsilon ) }{2}   \left( \bar{u}_{\text{b}} - \bar{u}_{\text{a}} + \frac{U_{\text{a}}^u}{U_{\text{a}}^r}\bar{r}_{\text{a}}\right)\frac{ U_{\text{a}}^u }{ U_{\text{a}}^r} + O(\epsilon) , \end{aligned} \nonumber \\
& \begin{aligned} & \bar{\varphi }_{\text{b}} =  \bar{\varphi }_{\text{a}}, && U_{\text{b}}^u = U_{\text{a}}^u,\end{aligned}  \nonumber \\
 & U_\text{b}^v = U_{\text{a}}^v- \epsilon  \log \left(1+ \frac{ \bar{u}_{\text{b}} - \bar{u}_{\text{a}}}{\bar{r}_{\text{a}}} \frac{ U_{\text{a}}^r}{ U_{\text{a}}^u} \right)U_{\text{a}}^u + O\left(\epsilon^2\right), \nonumber \\
 & U_\text{b}^r = U_{\text{a}}^r-\epsilon  \log \left(1+ \frac{ \bar{u}_{\text{b}} - \bar{u}_{\text{a}}}{\bar{r}_{\text{a}}} \frac{ U_{\text{a}}^r}{ U_{\text{a}}^u} \right) \frac{\left(U_{\text{a}}^u\right)^2}{U_{\text{a}}^r} + O\left(\epsilon^2\right).
\end{align}

As a result, we find that when $U_\text{a}^r \gg \epsilon$, the first-order correction to $\bar{v}_\text{b}$ and $\bar{r}_\text{b}$ scales as $\epsilon \log(\epsilon)$. Conversely, if $U_\text{a}^r \lesssim \epsilon$, the perturbation is at most linear in $\epsilon$. This means that a nonzero initial radial velocity amplifies the gravitational effects induced by the GW.

\section{Conclusions}\label{Conclusions}

In this work, we extended our previous results on the gravitational field of cylindrically shaped light pulses by deriving the full set of geodesics with zero angular velocity. We showed how the trajectory of a particle is perturbed when it interacts with the pulse or the accompanying GW. Then, we presented these results in the weak gravity limit ($\epsilon \ll 1$).

As an astrophysical application, we examined the case of gamma ray bursts, showing that they can accelerate an initially stationary test particle to velocities of up to $10^{-5}$ times the speed of light. Also, we found that the perturbations induced by the GW on a moving massive particle scale as $\epsilon \log(\epsilon)$, enhancing the effects previously predicted for particles initially at rest \cite{ylvn-3ybm}.

\section*{Acknowledgment}

We acknowledge financial support by the HORIZON-EIC-2022-PATHFINDERCHALLENGES-01 HEISINGBERG project 101114978.

\bibliographystyle{apsrev4-2}
\bibliography{bibliography}

\end{document}